\documentclass[nofootinbib,preprintnumbers]{article}
\usepackage{putex}
\usepackage{hyperref,epsfig}
\usepackage{graphicx}
\usepackage{adjustbox}
\usepackage[super]{nth}
\usepackage{epstopdf}
\usepackage{enumerate}
\usepackage{cite}
\usepackage{tensor}
\usepackage{slashed}

\numberwithin{equation}{section}

\newcommand {\be} {\begin {equation}}
\newcommand {\ee} {\end {equation}}

\newcommand {\bes} {\begin {equation*}}
\newcommand {\ees} {\end {equation*}}

\newcommand{\es}[2] {\begin{equation} \label{#1} \begin{split} #2 \end{split} \end{equation}}

\newcommand\Tstrut{\rule{0pt}{2.6ex}}         
\newcommand\Bstrut{\rule[-0.9ex]{0pt}{0pt}}   

\newcommand{\beq}{\begin{equation}}
\newcommand{\eeq}{\end{equation}}

\newcommand {\GeV} {\,\,\text{GeV}}

\def\Np{N_\text{pix}}

\begin{document}

\institution{PCTS}{Princeton Center for Theoretical Science, Princeton University, Princeton, NJ 08544}
\institution{PU}{Department of Physics, Princeton University, Princeton, NJ 08544}
\institution{MIT}{Center for Theoretical Physics, Massachusetts Institute of Technology, Cambridge, MA 02139}

\title{
Distinguishing Dark Matter from Unresolved Point Sources in the Inner Galaxy with Photon Statistics
}
\preprint{
MIT-CTP-4627}

\authors{Samuel K. Lee\worksat{\PCTS}, Mariangela Lisanti\worksat{\PU}, and Benjamin R.~Safdi\worksat{\MIT}
}

\abstract{
Data from the \emph{Fermi} Large Area Telescope suggests that there is an  extended excess of GeV gamma-ray photons in the Inner Galaxy.  Identifying potential astrophysical sources that contribute to this excess is an important step in verifying whether the signal originates from annihilating dark matter.  In this paper, we focus on the potential contribution of unresolved point sources, such as millisecond pulsars (MSPs).  We propose that the statistics of the photons---in particular, the flux probability density function (PDF) of the photon counts below the point-source detection threshold---can potentially distinguish between the dark-matter and point-source interpretations.  We calculate the flux PDF via the method of generating functions for these two models of the excess.  Working in the framework of Bayesian model comparison, we then demonstrate that the flux PDF can potentially provide evidence for an unresolved MSP-like point-source population.
}

\maketitle

\section{Introduction}

Weakly Interacting Massive Particles (WIMPs) are one of the leading candidates for dark matter.  While little is known about the particle nature of dark matter, there is strong evidence that it is concentrated at the Galactic Center (GC).  This high-density region provides an optimal environment for WIMPs to self-annihilate.  If the products of this annihilation are energetic Standard-Model final states, then observation of the photons that are produced in the ensuing decay of these states would constitute an indirect detection of WIMPs.  A detection of excess photons could then yield clues about the mass and annihilation rate of the WIMPs, thereby elucidating the particulars of the Standard-Model extension in which they are embedded.

A number of groups have indeed reported an excess of $\sim$GeV gamma rays at the GC and the Inner Galaxy ($\lesssim10^\circ$ from the GC)~\cite{0910.2998, 1010.2752, 1012.5839, 1110.0006, 1207.6047, 1302.6589, 1306.5725, 1307.6862, 1312.6671, 1402.4090, 1402.6703, 1410.6168, Murgia2014}.  The signal, which constitutes $\sim$10\% of the total flux in the Inner Galaxy, is found with high statistical significance.  It exhibits a spectrum that is compatible with a $\sim$35~GeV WIMP annihilating to $b$ quarks and has a morphology that is consistent with the expected dark-matter density profile.  However, this dark-matter interpretation may possibly be in tension with constraints from cosmic-ray~\cite{1404.3741, 1406.6027, 1407.2173, 1410.1527, 1410.6689} and radio~\cite{1406.6027, 1408.6224} observations, as well as \emph{Fermi} observations of dwarf galaxies~\cite{1310.0828, 1410.2242, Anderson2014}.  Thus, debate continues as to whether the excess can also be explained by standard astrophysical processes.  For example, although Refs.~\cite{1406.6948, 1409.0042, 1411.4647} find that the excess is robust across a variety of models for the diffuse gamma-ray background, systematic uncertainties in these models might nevertheless explain the excess. It could also be explained by injection of cosmic rays at the GC~\cite{1405.7685, 1405.7928} or unresolved point sources, such as young pulsars or millisecond pulsars (MSPs)~\cite{astro-ph/0501245, 1011.4275, 1305.0830, 1309.3428, 1402.4090, 1404.2318, 1406.2706, 1407.5625, 1411.2980, 1411.4363}.  

There are a number of reasons why the MSP explanation of the GeV gamma-ray excess is an intriguing possibility.  For one, the spectra of the MSPs observed by \emph{Fermi} appear to peak around 1--3~GeV, giving reason to suspect that an unresolved population of MSPs may be contributing to the excess~\cite{1011.4275, 1411.7410}.  Additionally, it is possible that the spatial distribution of MSPs in the Inner Galaxy is similar to that needed to explain the spatial morphology of the excess.  This latter piece of evidence is suggested by the observed spatial distribution of low-mass x-ray binaries (LMXBs) in M31~\cite{astro-ph/0610649, 1207.6047}; LMXBs are expected to be progenitors of MSPs.  On the other hand, several arguments against the MSP hypothesis have been raised~\cite{1305.0830, 1406.2706, 1407.5625}.  The MSP spectra are somewhat softer than that of the excess.  Furthermore, were there a population of MSPs in the Inner Galaxy large enough to produce the excess, it is unclear whether the number of resolved members of this population would exceed that detected by \emph{Fermi}~\cite{1305.4385}.  It is also unclear whether the number of LMXBs that might accompany this MSP population is already constrained by x-ray observations within the Milky Way.  Note, however, that these arguments rely quite heavily on \emph{Fermi} observations of the $\sim$60 identified MSPs~\cite{1305.4385}.  These may constitute a biased sample, have uncertain distance measures, and, being located primarily nearby and in the disk, may not be representative of a potentially different MSP population in the Inner Galaxy.

\begin{figure}[tb] 
\begin{center}
\includegraphics[trim=.1in 0in .1in 0in, clip=true, height=0.25\textwidth]{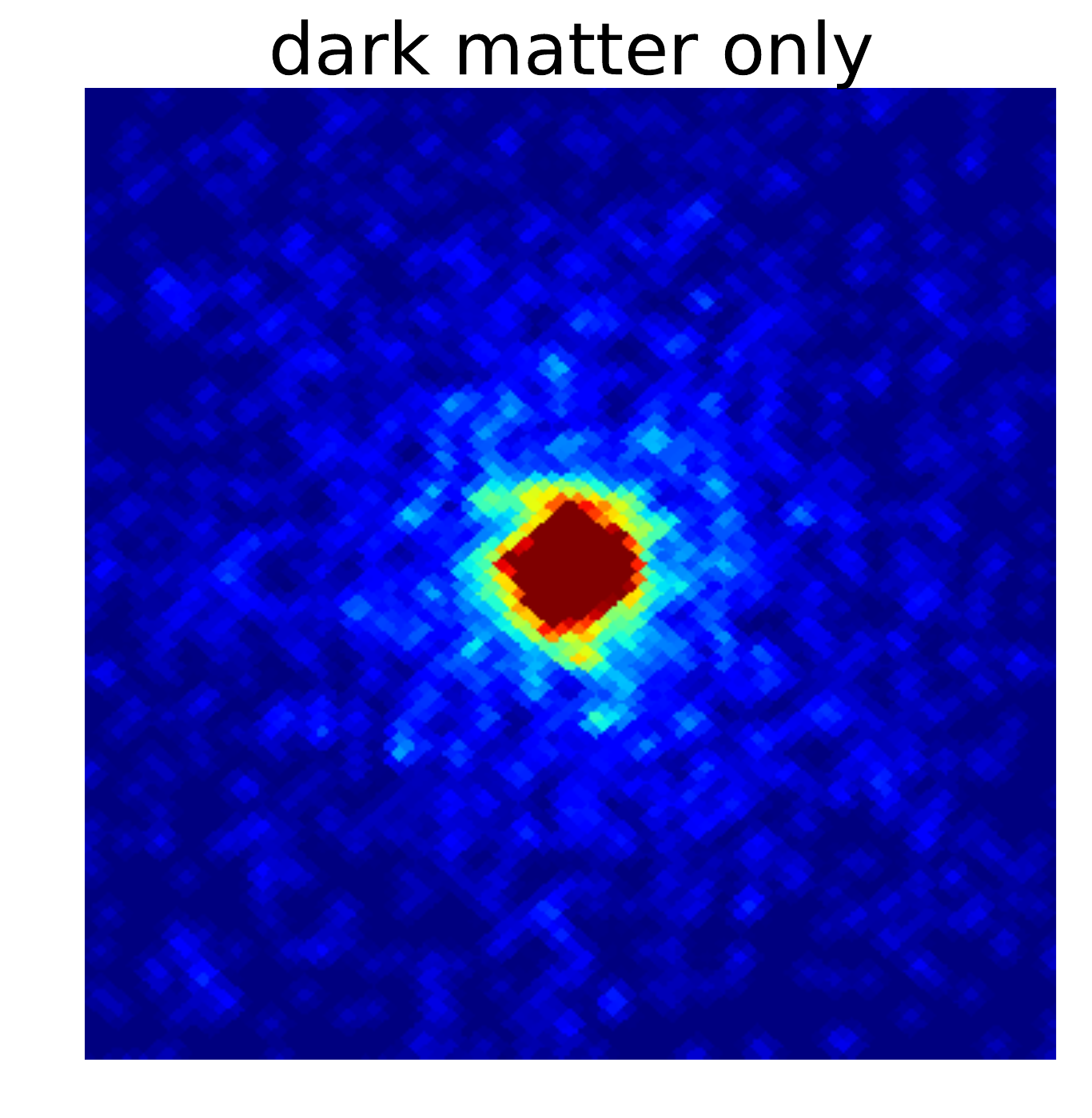}
\includegraphics[trim=.6in 0in .06in 0in, clip=true, height=0.25\textwidth]{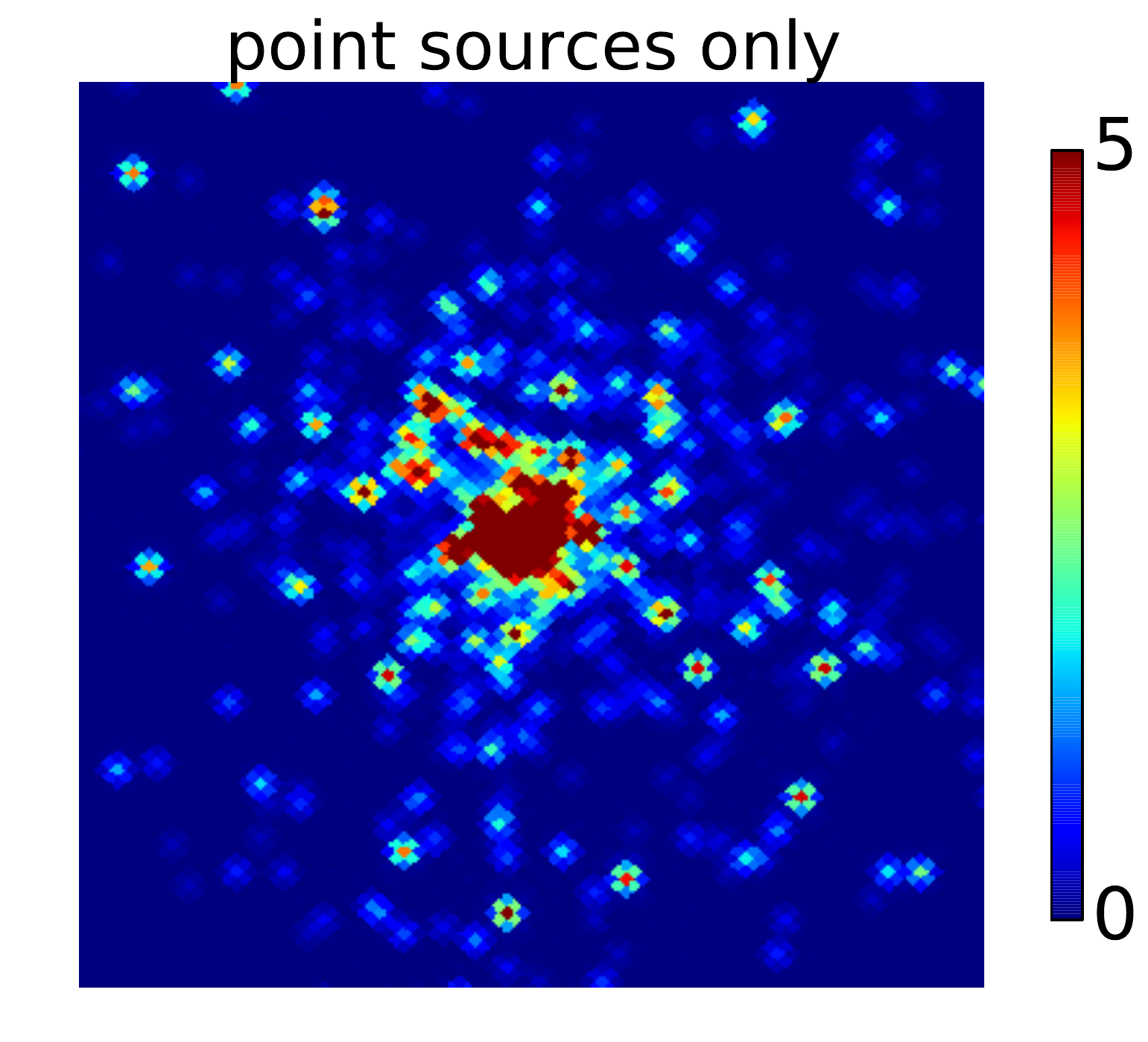}
\hspace{0.05in}
\includegraphics[trim=.6in 0in .1in 0in, clip=true, height=0.25\textwidth]{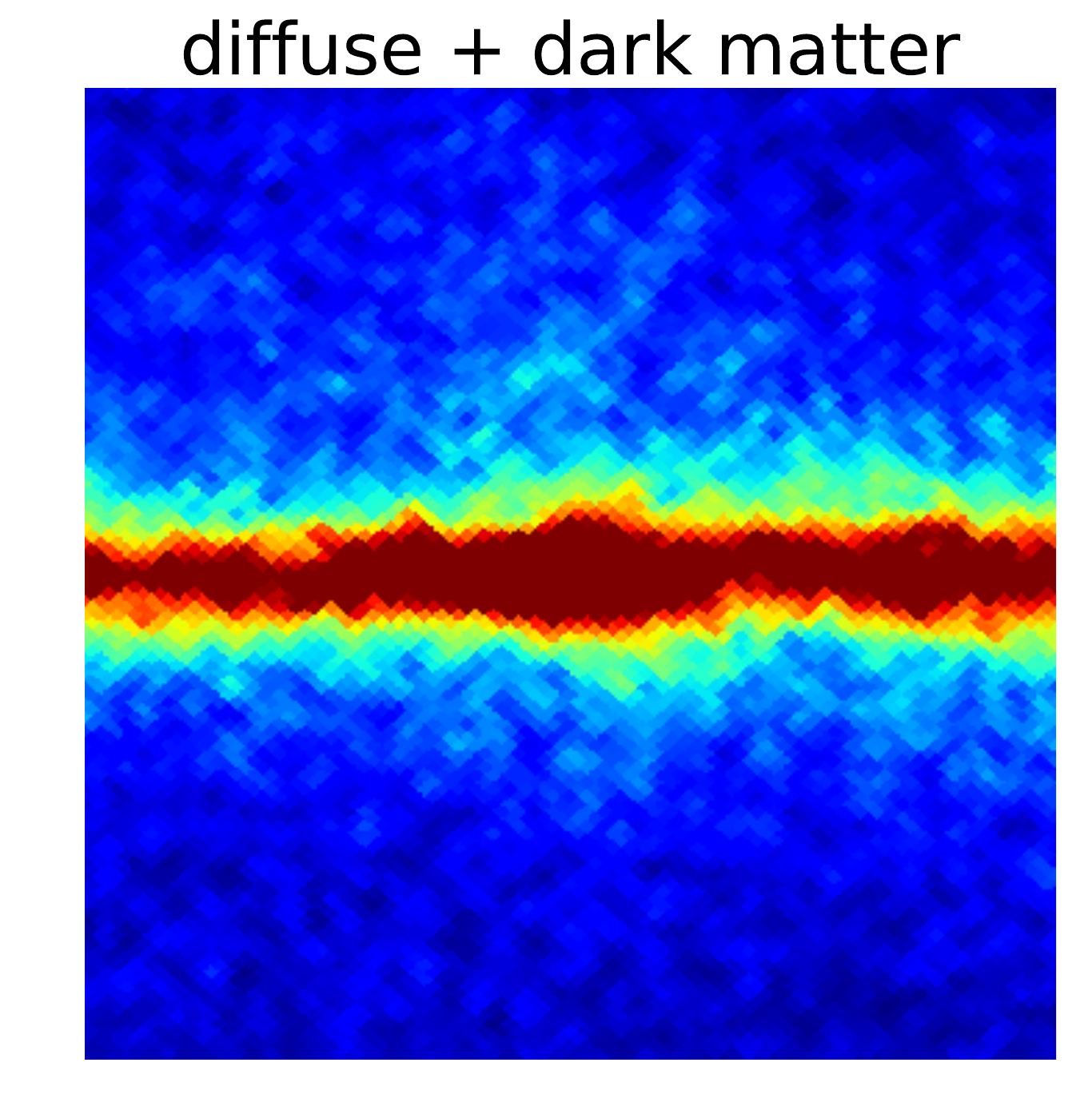}
\includegraphics[trim=.6in 0in .06in 0in, clip=true, height=0.25\textwidth]{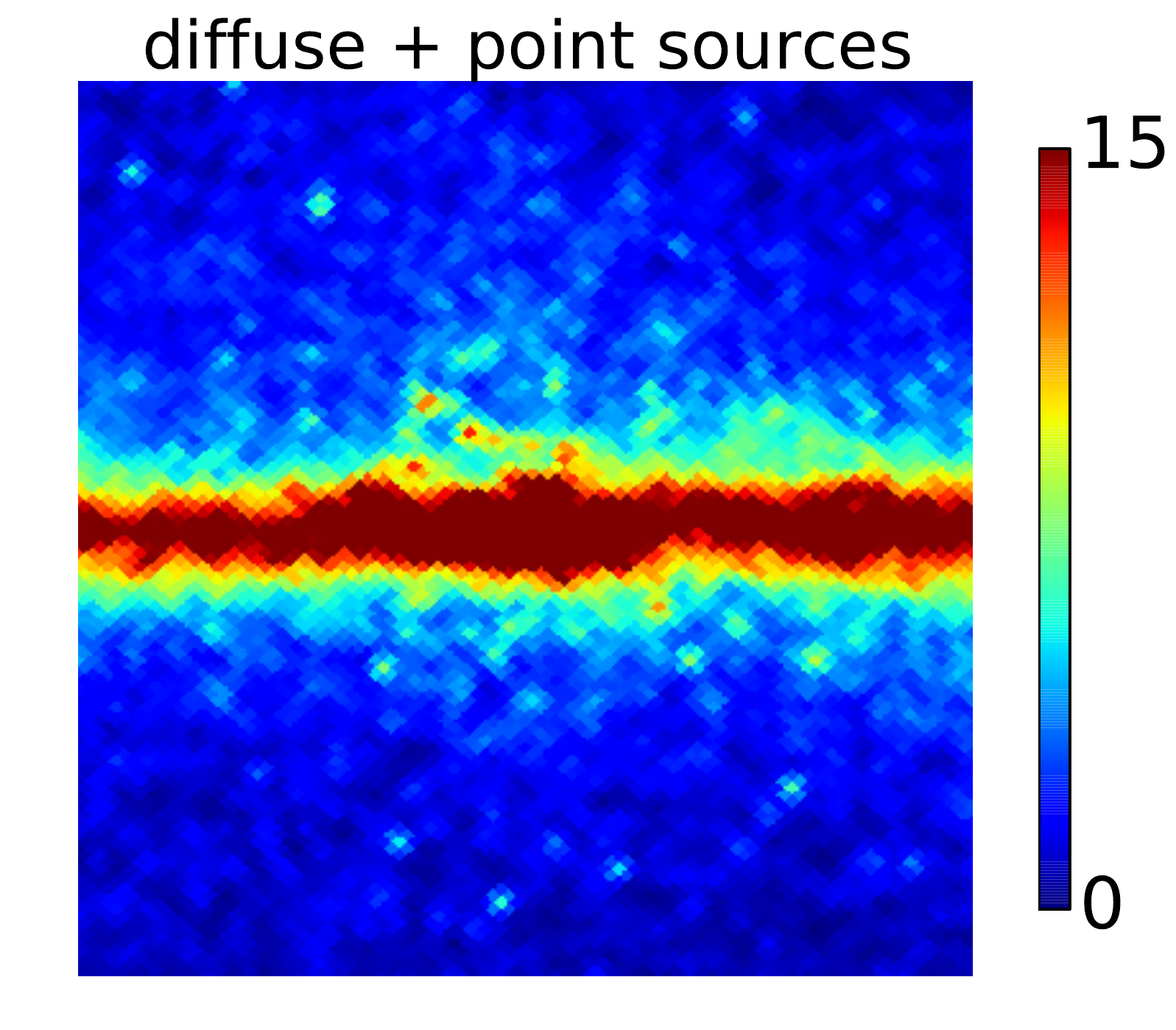}
\end{center}
\caption{\emph{(left panel)} Maps of simulated photon counts in a $20^\circ\times20^\circ$ region centered at the GC (clipped at 5 counts), for two scenarios consistent with the excess: 1) WIMP annihilation in the Galactic halo, and 2) MSP-like point sources.  These scenarios could potentially be distinguished by their photon statistics.  \emph{(right panel)} The same, but now including a simulated diffuse background (and clipping at 15 counts).  Since the excess comprises only $\sim$10$\%$ of the total flux in the Inner Galaxy, it is clear that a statistical analysis is required to test the point-source hypothesis.  All maps have been convolved with a $\sigma = 0.18^\circ$ Gaussian PSF and use a \textsc{HEALPix} pixelization with resolution parameter $nside = 256$.  The luminosity function taken for the point sources is explained in Sec.~\ref{sec:PS}, and has a maximum luminosity cutoff $L_\text{max} = 5 \times 10^{36}$~ph~s$^{-1}$. \vspace{-1em}}
\label{fig:PSF-maps}
\end{figure}

In principle, the presence of MSPs or other unresolved point sources can be inferred independently of the detection of individual sources.  This is possible because the statistics of the photon counts expected from a discrete population of point sources are fundamentally different from those expected from WIMP annihilation in the diffuse, smooth Galactic halo, as illustrated in Fig.~\ref{fig:PSF-maps}.  The left panel shows the photon counts (integrated from $\sim$2--12~GeV) in a $20^{\circ}\times20^{\circ}$ region centered at the GC for a simulated excess arising from 1) a 35~GeV WIMP annihilating to $b$ quarks and 2) MSP-like point sources.  There are clear similarities between the dark-matter and point-source excesses; they are both roughly spherically symmetric, for example.  However, the point-source excess gives rise to more pixels with either low or high photon counts.  Our goal is to capitalize on such differences in the photon-count statistics to distinguish the two scenarios.  However, the presence of the dominating diffuse background obscures these differences, as is demonstrated in the right panel of Fig.~\ref{fig:PSF-maps}.  Nevertheless, as we will show, a careful statistical analysis may still be able to distinguish the two scenarios.

In this work, we concentrate on the simplest photon-count statistic: the flux probability density function (PDF), or one-point function, which yields a histogram of the number of pixels with a given photon count.  The use of flux PDFs---sometimes referred to as ``fluctuation analysis" or ``P(D) analysis"---is standard in astronomical studies.  They have been used, for example, to identify active galactic nuclei (AGN) in the x-ray band~\cite{astro-ph/0111393} and star-forming galaxies in the infrared~\cite{1009.5675}.  This statistic has also been studied in the context of gamma-ray observations---for example, its ability to place limits on the presence of AGN~\cite{astro-ph/0201515,1104.0010}, MSPs~\cite{0904.3102}, and dark-matter subhalos~\cite{0810.1284, 1006.2399} at high Galactic latitudes, where the diffuse background is less dominant, has been examined.\footnote{The use of a similar statistic to constrain point-source populations in gamma-ray observations was also discussed in Ref.~\cite{0910.0482}.}

By focusing on the flux PDF, our analysis is largely independent of the detailed astrophysics of the point sources.  That is, while we are motivated by the MSP explanation of the excess, the point-source population may be comprised of other astrophysical compact objects.  Furthermore, unresolved extended structures---such as dark-matter subhalos, molecular clouds, etc.---could also be possible ``point-source'' candidates.  The tests we propose will be equally applicable to all of these populations.

Our paper presents the first detailed study of the viability of the flux PDF to distinguish unresolved gamma-ray point sources in the Inner Galaxy.  In Sec.~\ref{sec:photonstats}, we describe a general formalism for the calculation of the flux PDF for emission from both diffuse components and point sources.  We next give specific dark-matter and point-source models of the excess in Sec.~\ref{sec:models}, calculating the flux PDFs for these models and verifying our results with simulated data.  Then, in Sec.~\ref{sec:Bayesian}, we use Bayesian model comparison to show that over a wide range of well-motivated point-source luminosity functions, our procedure is able to distinguish dark-matter and point-source explanations of the excess.  Finally, in Sec.~\ref{sec:discussion}, we examine issues that may arise in the application of our procedure to the real data and discuss implications for future observations and analyses.

\section{The statistics of photon counts}
\label{sec:photonstats}

The aim of this section is to construct a photon-count statistic that, even in the presence of a dominating diffuse background, can be used to distinguish between two different hypotheses for the excess: 1) WIMP annihilation in the smooth, diffuse Galactic halo, and 2) emission from unresolved point sources.  One such statistic is the flux PDF, which has long been used in astronomical studies to detect the presence of unresolved point sources.  The flux PDF is often called a $P(D)$ distribution, as it was first used for observations of faint radio sources that produced deflections of the measuring apparatus~\cite{1957PCPS...53..764S}, and it is usually computed using characteristic functions (see also, for example, Refs.~\cite{1974ApJ...188..279C,astro-ph/9801293,astro-ph/0201515}).  In this work, we follow the more elegant derivation in Ref.~\cite{1104.0010}, which employs the method of generating functions.

Consider a region of interest (ROI) on the sky that contains $\Np$ pixels.\footnote{We use the standard \textsc{HEALPix} pixelization of the sphere~\cite{astro-ph/0409513}, which is specified by the resolution parameter $nside$.  The number of pixels over the full sky for a particular $nside$ is then $12 \times nside^2$.}  Suppose that $n_k$ of these pixels contain $k$ photons after an observation is made.  Then, the quantity 
\es{Pk}{
\hat{p}_k = {n_k \over \Np} 
}
estimates the probability $p_k$ of finding $k$ photons per pixel, averaged over all pixels in the ROI.  We refer to $p_k$ as the ``flux'' PDF throughout, although strictly speaking it is the PDF for the photon counts.

The goal of the following subsections is to obtain the flux PDF $p_k$ in the ROI that appropriately accounts for all components of the emission.  This flux PDF is specified by its generating function.  In general, the generating function $\mathcal{P}(t)$ for a non-negative, integer-valued random variable $k$ is defined as a power series in an auxiliary variable $t$ as follows:
\es{GenFuc}{
\mathcal{P}(t) = \sum\limits_{k=0}^\infty p_k t^k \, .
}
Derivatives of the generating function with respect to $t$ yield  the flux PDF
\es{GenFuc}{
p_k = \frac{1}{k!} \frac{d^k\mathcal{P}}{dt^k}\bigg|_{t=0} \,.
}
Because the generating function for the sum of independent random discrete variables is given by the product
of the generating functions for those variables, the total flux PDF follows simply once the flux PDFs of the individual components of the emission are determined.  Therefore, we begin in Sec.~\ref{sec:genfunc} by finding the generating functions for the flux PDFs of uniform and non-uniform diffuse emission, as well as point-source emission.  Once the generating functions for these different components are obtained, corrections due to the point-spread function must be included (Sec.~\ref{sec:stat_PSF}).  
We may then write down a generating function for the flux PDF that models the photon counts from all components of the emission in the ROI (Sec.~\ref{sec:stat_InnerGalaxy}).

\subsection{Generating Functions}
\label{sec:genfunc}

\subsubsection{Uniform diffuse emission}
\label{sec:stat_iso}

We begin by considering the case where the photons arise from diffuse emission that is uniform in the ROI (\emph{i.e.}, isotropic emission, if the ROI covers the full sky).  The probability of finding $k$ photons in a given pixel is given by the Poisson distribution
\es{istrop}{
p_k =  {x^k  \over k!}e^{-x}\,,
}
where $x$ is the mean number of photons expected in the pixel.   In this case, the mean photon count per pixel is uniform and does not depend on the pixel position; it is given by $x_\text{iso} = N_\gamma / \Np$, where $N_\gamma$ is the expected number of observed photons.  The average generating function for the ROI is 
\es{I_gen}{
\sum\limits_{k=0}^\infty p_k t^k = \exp\left[x_\text{iso} (t-1) \right] \equiv I(t)\,.
}
This generating function indeed yields the Poisson distribution, as can be easily confirmed using Eqs.~\eqref{GenFuc} and \eqref{I_gen}.

\subsubsection{Non-uniform diffuse emission}
\label{sec:stat_var}
The Galactic diffuse background and WIMP annihilation both result in spatially varying, non-uniform diffuse emission.  For non-uniform diffuse emission, the mean photon count per pixel depends on the position of the pixel, so we denote it as $x_\textrm{var}^p$, where the superscript labels the pixel number.  By averaging over the individual flux PDFs in each pixel, we can find the average flux PDF over the ROI.  This is obtained by averaging the Poisson generating function for each pixel:  
\es{G_gen}{
\sum\limits_{k=0}^\infty p_k t^k = \frac{1}{\Np}\sum_{p=1}^{\Np} \exp\left[x_\text{var}^p (t-1)\right] \equiv G(t)\,.
}
The net result is a flux PDF that is more complicated than a simple Poissonian. 

\subsubsection{Point-source emission}
\label{sec:stat_PS}

Finally, if the photons are emitted from a population of point sources, then the flux PDF in the ROI does not follow a Poisson distribution.  Intuitively, the $p_k$ should be larger at both small and large $k$, relative to the case of uniform diffuse emission.  The larger values of $p_k$ at small $k$ are due to the empty or faint pixels between bright sources, while those at large $k$ result from the fact that some pixels contain individual bright sources or multiple sources that are cumulatively bright.  The derivation of the $p_k$ for point sources is detailed in Appendix A of Ref.~\cite{1104.0010}, and we simply state the result here.  Let $x_m$ be the mean number of sources per pixel in the ROI that contribute $m$ observed photons to the pixel that contains them.  The generating function is then given by
\es{P_gen}{
\sum\limits_{k=0}^\infty p_k t^k =\exp\left[\sum_{m=1}^{\infty} x_m (t^m - 1) \right] \equiv P(t)\,.
}
Notice that the generating function for uniform diffuse emission given by Eq.~\eqref{I_gen} is equivalent to that for a population of ``1-photon'' point sources (\emph{i.e.}, one with nonzero $x_1$ and all other $x_m$ vanishing).

The generating function for point-source emission is thus determined by the mean number of $m$-photon sources per pixel, $x_m$.  The value of $x_m$ depends on the source-count distribution $dN / dS$, which gives the total number of sources $N$ in the ROI that individually contribute an average of $S$ photon counts: 
\es{xm}{
x_m = {1 \over \Np} \int\! dS \, {dN \over dS} {S^m \over m!} e^{-S} \,.
}
Here, we assume that if a point source contributes an average of $S$ counts, then the probability of observing $m$ photons from that source is given by the Poisson distribution with mean $S$.

As in Ref.~\cite{1104.0010}, we consider source-count distributions that are modeled by a broken power law, which is specified by four parameters---the normalization $A$, the location of the break $S_b$, and the indices $n_1$ and $n_2$ above and below the break, respectively:
\es{dNdS}{
{dN \over dS} = A
\left\{ \begin{array}{cc}
\left({S\over S_b}\right)^{-n_1} \,, & S \geq S_b \,, \\
\left({S\over S_b}\right)^{-n_2} \,, & S < S_b \,. \\
\end{array} \right.
}
We shall consider cases where $n_1 > 2$ and $n_2 < 2$, for which this distribution yields non-divergent total source counts and fluxes.  We may then exactly integrate Eq.~\eqref{xm}:
\es{xmA}{
x_m = \frac{A}{\Np\, m!} \left[ S_b^{n_1} \Gamma(1 - n_1 + m, S_b) + S_b^{n_2} \left(\Gamma(1 - n_2 + m) - \Gamma(1 - n_2 + m, S_b)\right)\right]  \,,\,\, \, \, m \geq 1\,.
}

\subsection{Point-spread function corrections}
\label{sec:stat_PSF}

The expression in Eq.~\eqref{xmA} for the mean number of $m$-photon sources per pixel is valid in the limit of a vanishing point-spread function (PSF).  A finite PSF redistributes the flux from a point source over multiple pixels.  The dependence of the flux PDF on the PSF thus depends on both the properties of the PSF and the chosen pixelization.  To quantify this dependence, we define the distribution $\rho(f)$, where $ \rho(f) df$ is the number of pixels that observe a fraction between $f$ and $f + df$ of the flux from a single point source.  The distribution $\rho(f)$ is normalized as $\int_0^1\! df f \rho(f) = 1$.  As the angular scale of the PSF decreases, the function $\rho(f)$ approaches a Dirac $\delta$-function at $f = 1$.

The function $\rho(f)$ can be determined from Monte Carlo simulations  following the procedure outlined in Ref.~\cite{1104.0010}.  To summarize, a Gaussian distribution with standard deviation $\sigma$ is randomly placed on a pixelated sphere, and $1,000$ points are drawn from it.  From this, one can determine the fraction of points $f_i$ that fall within the $i^\text{th}$ pixel.  The function $\rho(f)$ is  approximated as
\es{rhoF}{
\rho(f) = {\Delta n(f) \over \Delta f} \,,
}
where 
$\Delta n(f)$ is the number of pixels with fractions between $f$ and $f + \Delta f$.  This process is repeated for $50,000$ Gaussian distributions, each placed randomly on the sphere, and the individual approximations to the function $\rho(f)$ are averaged together.  

\begin{figure}[tb] 
\begin{center}
\includegraphics[trim=0in 0in 0in 0in, clip=true, width=0.6\textwidth]{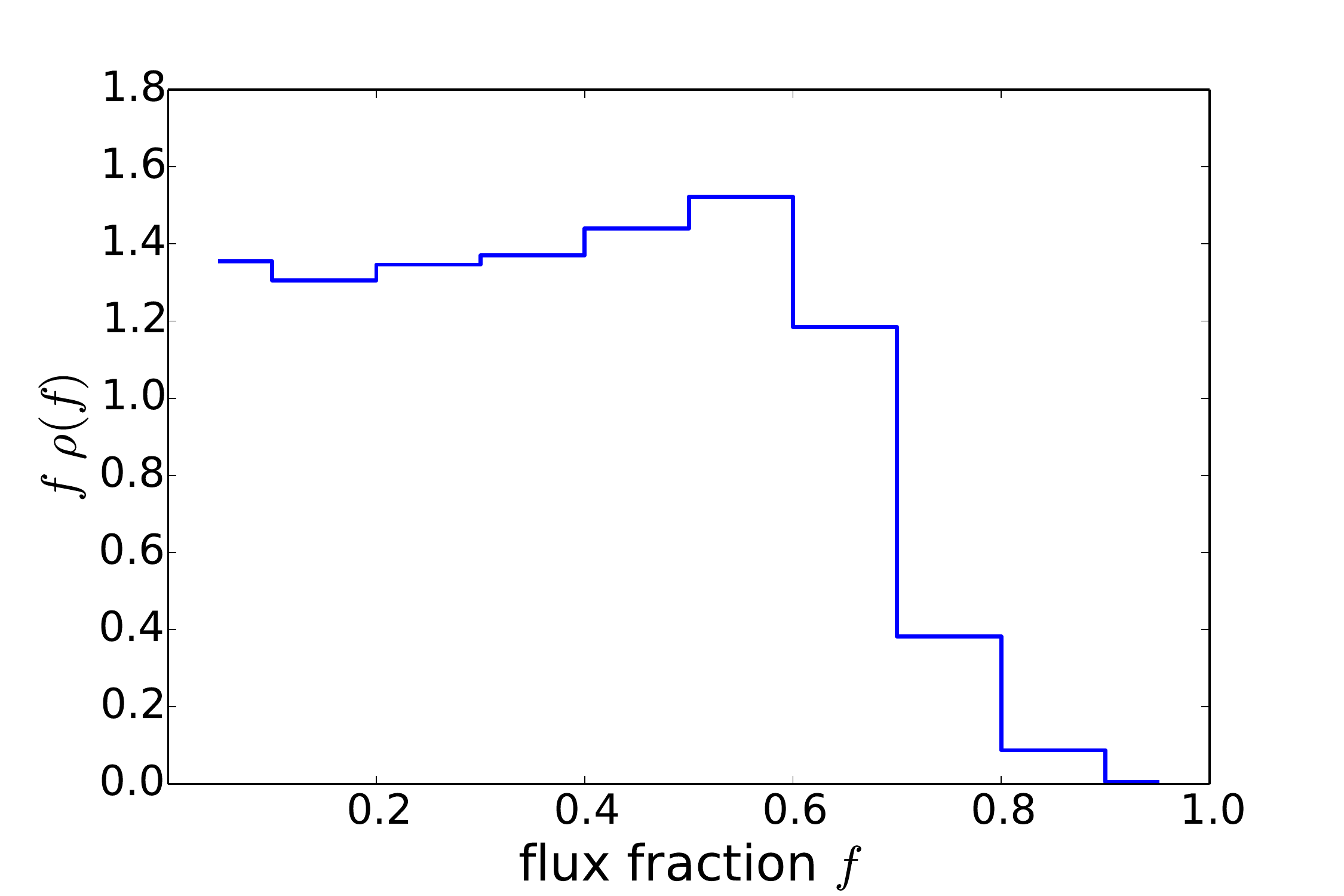}
\end{center}
\caption{The average distribution across all pixels of the fraction $f$ of the flux from a single point source.  As a result of the pixelization of the sphere, as well as the finite point-spread function, most pixels contain less than $\sim$$60\%$ of the flux from the point source.}
\label{fig:PSF}
\end{figure}

The resulting $f \rho(f)$ is plotted in Fig.~\ref{fig:PSF}, which shows the average distribution across all pixels of the fraction $f$ of the flux from a single point source.  We use a \textsc{HEALPix} parameter $nside = 128$, which gives pixels $\sim$$0.46^\circ$ to a side.  In addition, we assume an energy-averaged Gaussian PSF with $\sigma = 0.18^\circ$.  This is obtained by weighing the energy-dependent PSF for events with a cut on the event-reconstruction parameter CTBCORE\footnote{Throughout this work, we assume a Gaussian PSF and energy-binned exposures roughly consistent with a Q2 cut (top half) on the CTBCORE parameter.  Note, however, that the PSF resulting from such a cut is not exactly Gaussian.  See Ref.~\cite{1406.0507} for details.}  by the observed energy spectrum for the excess in the energy range 1.9--11.9~GeV.    As the figure illustrates, the PSF distributes photons from an individual point source across pixels such that most of the pixels contain less than $\sim$60\% of the total flux.

Given the function $\rho(f)$, one can include the effect of the PSF in the calculation of the mean number of $m$-photon sources per pixel.  The new result is a modification of Eq.~\eqref{xm} to
\es{xmPSF}{
x_m &=  {1 \over \Np}  \int_0^1\! df\, \rho(f) \int\! dS \, {dN \over dS}  {(f \, S)^m \over m!} e^{-f \,S} \\
&= \int_0^1\! df\, \frac{\rho(f)}{f}\, \tilde x_m(f) \,,
}       
with the $\tilde x_m(f)$ given by Eq.~\eqref{xmA} with $S_b \to f S_b$.  By combining this with Eq.~\eqref{P_gen}, we obtain the finite-PSF point-source generating function $P(t)$.

\subsection{The total flux PDF for the Inner Galaxy}
\label{sec:stat_InnerGalaxy}

The functions $I(t)$, $G(t)$, and $P(t)$ are generating functions associated with the flux PDFs for pure uniform-diffuse, pure non-uniform--diffuse, and pure point-source components of the gamma-ray sky, respectively.  In reality, the Inner Galaxy contains contributions from each of these source components.  Therefore, the generating function for the total flux PDF is simply the product of the generating functions for each component:
\es{nuGen}{
\sum_{k=0}^\infty p_k (x_\text{iso}, x_\text{var}^p, \alpha) t^k = I(t; x_\text{iso})\, G(t; x_\text{var}^p)\, P(t; \alpha) \,.
}
As suggested by the notation, the flux PDF depends on the mean uniform-diffuse counts $x_\text{iso}$, the mean counts ${x_\text{var}^p = x_\text{var,bg}^p + x_\text{var,DM}^p}$ for the spatially varying diffuse-background and dark-matter components, and the set of point-source parameters ${\alpha = \{A, S_b, n_1, n_2\}}$.  The values $x_\text{var,bg}^p$ may be computed from a given model of the Galactic diffuse emission and other components of the diffuse background.  However, there is a degeneracy between $x_\text{iso}$ and the uniform part of the $x_\text{var}^p$; changing the value of $x_\text{iso}$ is equivalent to shifting all of the $x_\text{var}^p$ by a constant.  We shall therefore absorb the parameter $x_\text{iso}$ into the background counts $x_\text{var,bg}^p$, so that the $x_\text{var}^p$ completely determine the diffuse emission and the remaining four parameters ${\alpha = \{A, S_b, n_1, n_2\}}$ determine the emission from point sources.  The factor of $I(t; x_\text{iso})$ in Eq.~\eqref{nuGen} can be removed accordingly.

We treat the shifted background counts $x_\text{var,bg}^p$ as known and fixed parameters.  This approach is compatible with the template-based analyses that have studied the excess~\cite{1110.0006, 1207.6047, 1302.6589, 1306.5725, 1307.6862, 1312.6671, 1402.4090, 1402.6703, 1410.6168, Murgia2014}.  That is, these analyses recover an excess of photons with respect to a given spatial template for the background.  The energy spectrum of these photons is then used to argue that the excess arises from a component that is different from the background.  Our goal here is to attempt to go a step further, by using the \emph{statistics} of these photons to determine if the excess is more consistent with diffuse emission from WIMP annihilation or emission from point sources.

\section{Dark-matter and point-source models of the excess}
\label{sec:models}

Next, we consider simple models of the gamma-ray sky with only two components.  The first component is the excess; we shall consider two possibilities for its origin: 1) WIMP annihilation in the Galactic halo, or 2) MSP-like point sources.  The second component is the diffuse gamma-ray background, which is dominated in the Inner Galaxy by Galactic emission from inverse-Compton, bremsstrahlung, and $\pi^0$-decay subcomponents.  We neglect contributions from large-scale structures (\emph{e.g.}, the Fermi Bubbles~\cite{0910.4583, 1005.5480, 1407.7905}) and from other point-source populations (\emph{e.g.}, AGN), as these are expected to be sub-dominant sources of photons in the ROI we shall consider.  For example, unresolved AGN contribute $\sim$$20$\% of the flux at high latitudes~\cite{1104.0010}, but only $\sim$1--2\% of the total flux in the Inner Galaxy.

We now discuss the dark-matter, point-source, and Galactic-diffuse components in turn, demonstrating how to calculate their corresponding flux PDFs using the formalism outlined in Sec.~\ref{sec:photonstats}.  Before proceeding, we briefly note a number of assumptions that we take throughout the analysis.  First, we assume an exposure of $\sim$$4\times 10^{10}$~cm$^2$~s and a Gaussian PSF with $\sigma = 0.18^\circ$, which are roughly consistent with a Q2 cut on the CTBCORE parameter~\cite{1406.0507}.  Second, we conservatively limit the flux-PDF analysis to the ROI defined by $5^\circ \leq \psi \leq 10^\circ$ and $|b| \geq 2.5^\circ$, where $\psi$ is the angle from the GC and $b$ is the Galactic latitude; this is to reduce contamination from the diffuse emission in the Galactic plane and to avoid the complexities of the GC.  This ROI contains 882 pixels in the $nside = 128$ \textsc{HEALPix} pixelization we use throughout.  Third, we only focus on photons in the energy range 1.9--11.9~GeV.  The reason for this is that the energy spectrum of the Galactic emission approximately falls as a power law, so the flux is dominated by the diffuse background at lower energies.  Meanwhile, the spectrum of the excess peaks around 1--3~GeV and extends up to $\sim$10~GeV.  We thus select the range from 1.9--11.9~GeV to strike a balance between maximizing the signal-to-noise ratio of the excess to the background and maintaining a sufficient number of photons.

 \subsection{Dark-matter excess}
\label{sec:DM}

Let us begin by considering the hypothesis that the excess arises from dark-matter annihilation.  If the annihilation in the Inner Galaxy is dominated by a smooth, spatially varying component, the associated flux PDF is given by the generating function $G(t)$ for non-uniform diffuse emission (see Eq.~\eqref{G_gen}).  The function $G(t)$ depends on the mean photon counts per pixel, $x^p_\text{var,DM}$, which are determined by the intensity profile for a given dark-matter--annihilation model.

This intensity profile depends on the halo density profile $\rho(r)$, as well as the dark-matter mass $m_\chi$ and annihilation cross section (times relative velocity) $\sigma v$.  The differential number intensity of photons with energy $E$ at an angle $\psi$ from the GC is 
\es{LOSDM}{
\Phi (E, \psi) = \frac{\sigma v }{8\pi m_\chi^2} \frac{dN_\gamma}{dE} \int\! d\ell\, \rho[r(\ell, \psi)]^2 \propto \psi^{1 - 2 \gamma} \,,
 }
 where $\ell$ is the line-of-sight (LOS) distance from Earth and $dN_\gamma/dE$ is the energy spectrum of the photons.  We assume a generalized Navarro-Frenk-White halo~\cite{astro-ph/9508025, astro-ph/9611107} with density profile
\es{gNFW}{
\rho(r) = \rho_\ast {(r/r_s)^{-\gamma} \over (1 + r/r_s)^{3-\gamma} } \,,
}
 where $r$ is the distance from the GC, $\gamma$ is the inner slope, $r_\text{S} = 20$~kpc, and $\rho_\ast$ is set such that the density is $0.3\GeV$~cm$^{-3}$ at $r_0 = 8.5$~kpc.  For this density profile, the intensity scales roughly as $\psi^{1-2\gamma}$ for small $\psi$; for the best-fit value $\gamma = 1.26$ found by Ref.~\cite{1402.6703}, it falls as $\psi^{-1.5}$.  Ref.~\cite{1402.6703} also found that the excess is best-fit by $m_\chi \approx  35\GeV$ WIMPs annihilating into $b$ quarks with an annihilation cross section $\sigma v \approx 1.7 \times 10^{-26}$~cm$^3$~s$^{-1}$.\footnote{Different masses and decay channels are also consistent with the data, especially once background systematics~\cite{1409.0042, 1411.2592, 1411.4647} or secondary emission~\cite{1307.7152, 1403.1987} are accounted for. }  Thus, we henceforth assume these parameters,  obtaining the photon energy spectrum for annihilation to $b$ quarks from \textsc{Pythia}~\cite{0710.3820}, which yields $\sim$1.7 photons per annihilation in the range 1.9--11.9~GeV.  Assuming this best-fit dark-matter--annihilation model, the excess contributes $\sim$3500 photons that are found within $\psi \lesssim 10^\circ$, with $\sim$650 of these photons falling within the ROI.

\begin{figure}[p]
\makebox[\linewidth]{
\includegraphics[trim=.5in 0in 0in 0in, clip=true, height=0.3\textwidth]{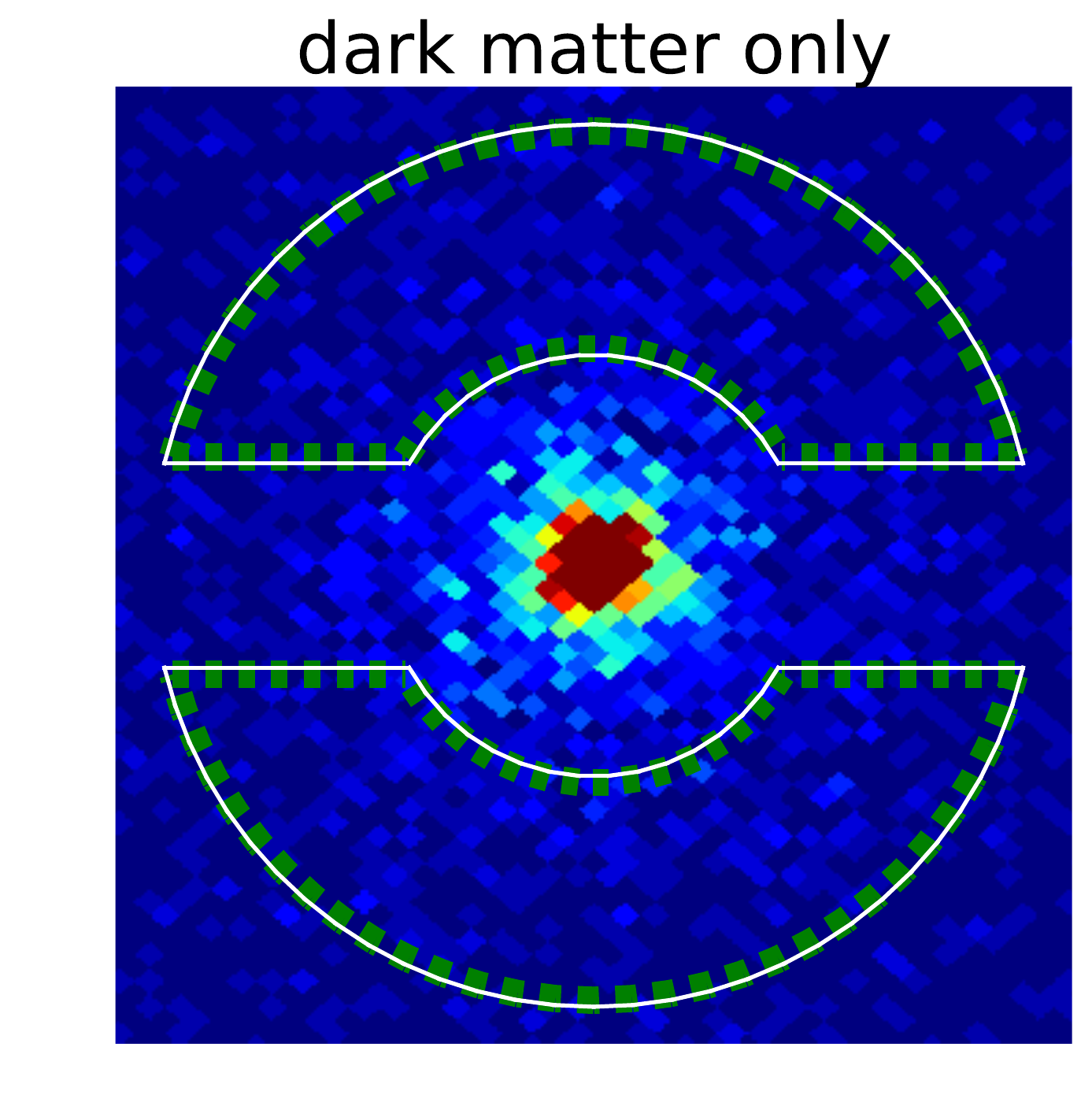}
\includegraphics[trim=.5in 0in 0in 0in, clip=true, height=0.3\textwidth]{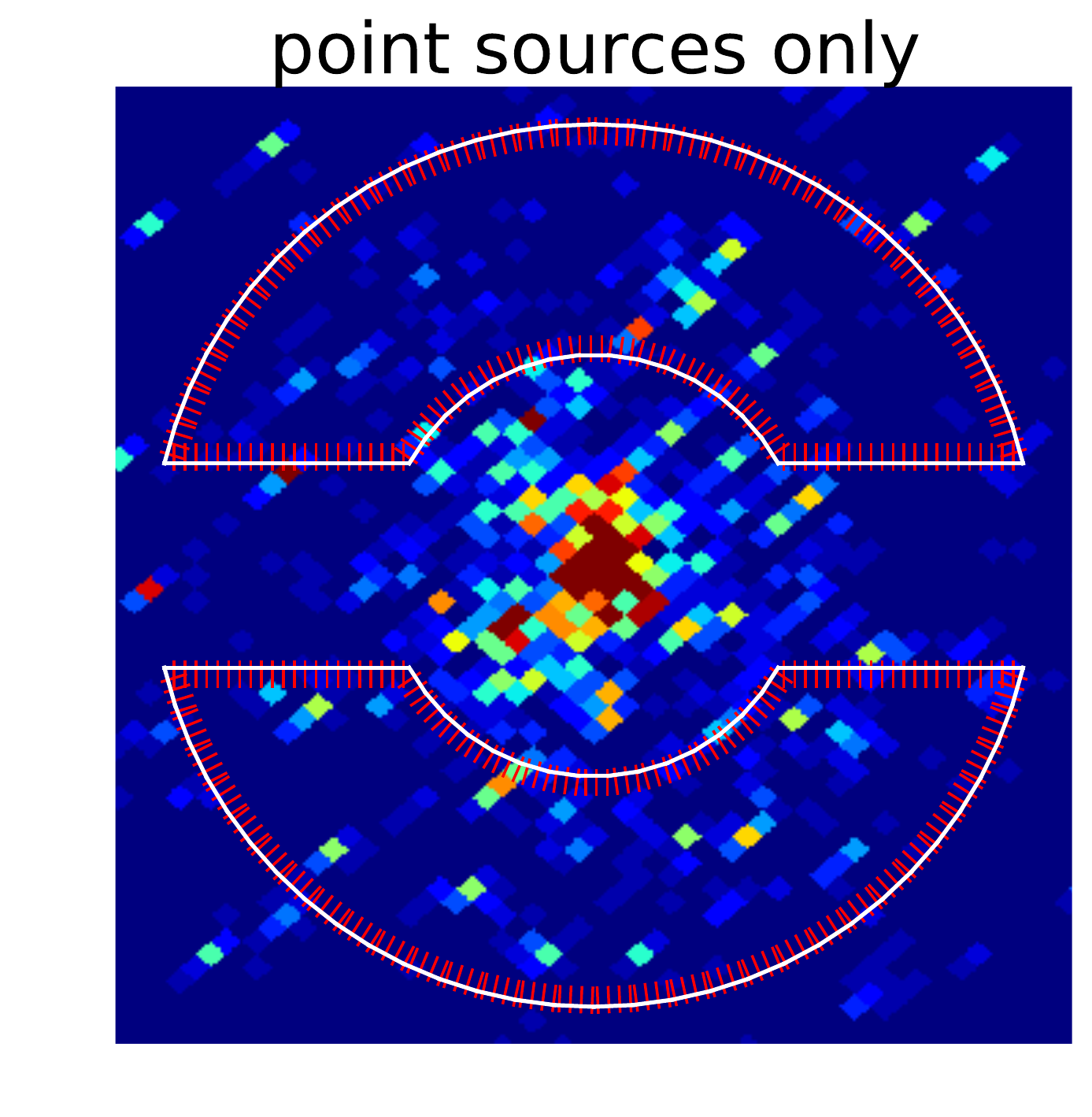} 
\includegraphics[trim=.5in 0in .08in 0in, clip=true, height=0.3\textwidth]{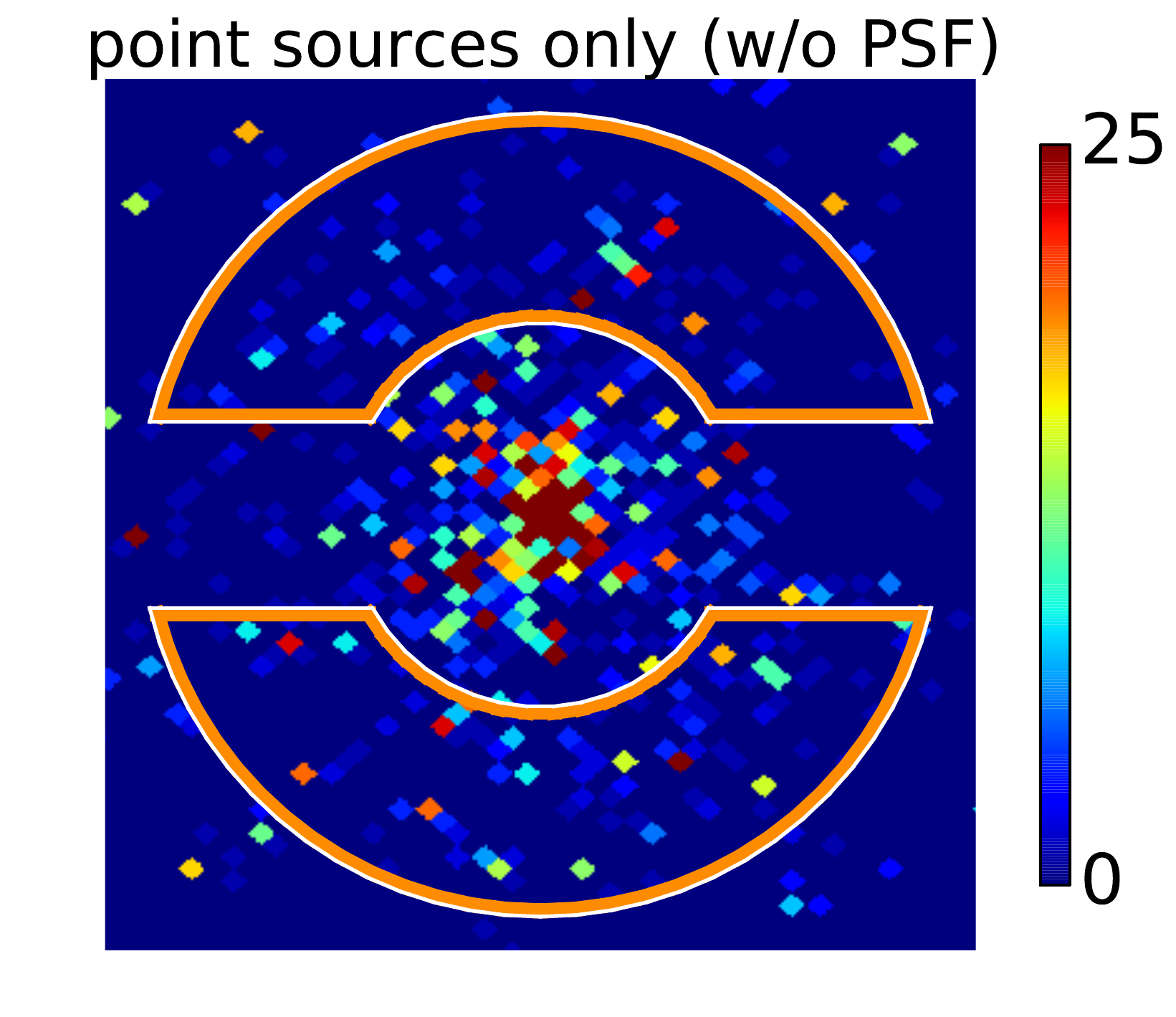} 
}
\caption{Maps of simulated photon counts in the Inner Galaxy, assuming that the GeV excess results from \emph{(left)} WIMP annihilation in the Galactic halo, \emph{(center)} MSP-like point sources (including the effect of a $\sigma = 0.18^\circ$ Gaussian PSF), and \emph{(right)} MSP-like point sources (neglecting the PSF).  Only photons in the energy range from 1.9--11.9~GeV are shown, and counts are clipped at 25.  The ROI, defined by $5^\circ \leq \psi \leq 10^\circ$ and $|b| \geq 2.5^\circ$, is indicated in each map.   Note that an excess comprised of point sources results in a larger fraction of pixels having either zero or a high number of counts (the latter corresponding to individual bright sources); however, the number of pixels with high counts is reduced by the PSF.  For the point sources, a maximum luminosity cutoff $L_\text{max} = 5 \times 10^{36}$~ph~s$^{-1}$ is assumed.}
\label{fig:maps-no-diff}
\vspace{0.5in}
\makebox[\linewidth]{
\includegraphics[trim=0in 0in 0in .5in, clip=true, width=0.6\textwidth]{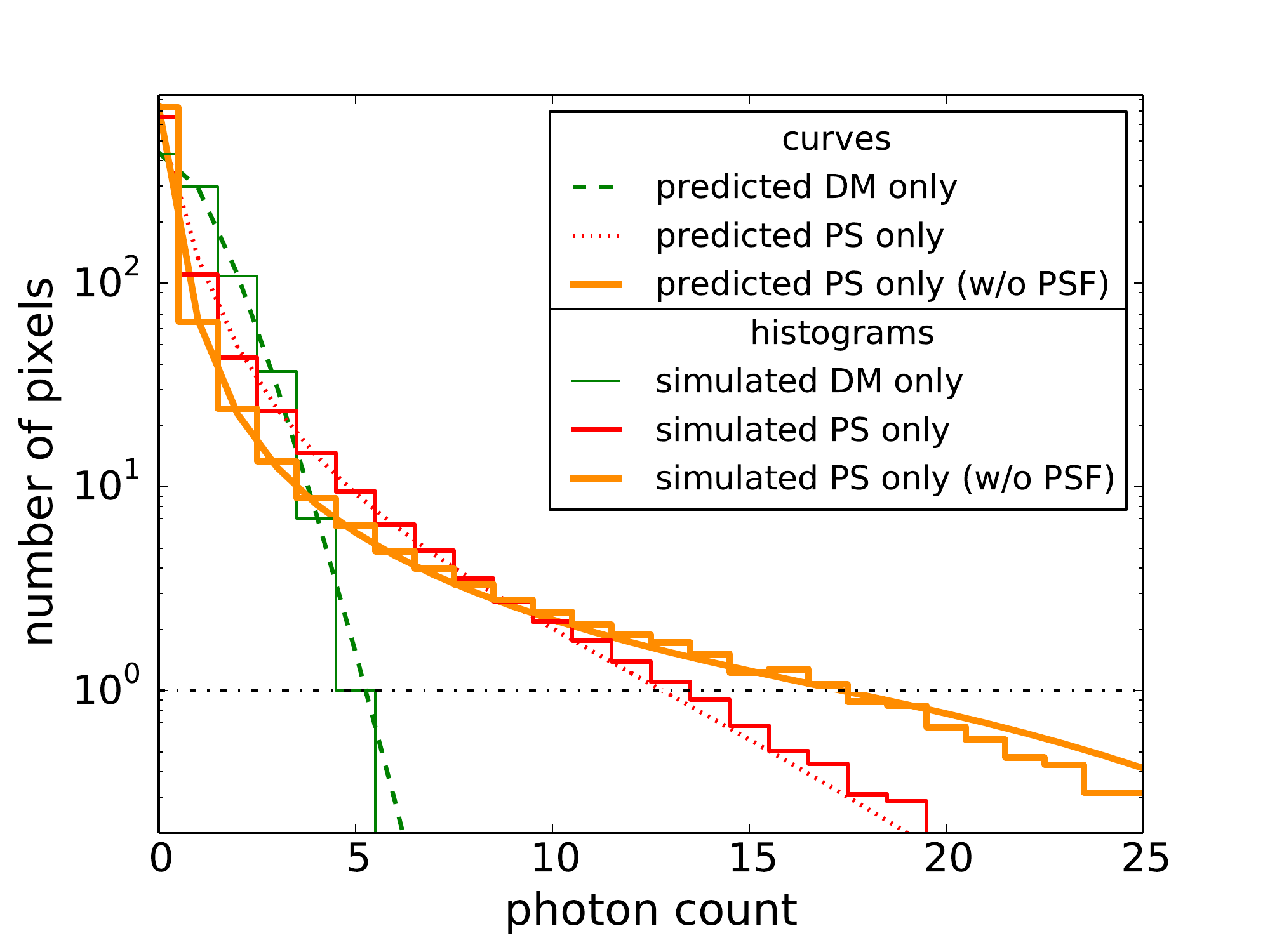}
}
\caption{The average flux PDFs in the ROI (normalized to the number of $nside = 128$ \textsc{HEALPix} pixels contained there) for dark matter (DM, green) and point sources (PS) with (red) and without (orange) the PSF.  The curves indicate the predicted PDFs, which are obtained by applying the formalism of generating functions.  The histograms are constructed by averaging the PDFs from 500 simulated data sets.  There is good agreement between the predicted and simulated PDFs.  Furthermore, the difference between the nearly Poissonian PDF of the dark-matter excess and the non-Poissonian power-law PDF of the point sources is clear, although diminished by the effect of the PSF.  The fact that point sources give a larger fraction of pixels with either zero or a high number of counts is also evident.  The dash-dotted--black line indicates where the number of pixels falls below one; the predicted and simulated PDFs plotted here cross below this line, and hence yield fractional pixel numbers, but the PDF of a real data set is limited to integer pixel numbers.}
\label{fig:PDF-no-diff}
\end{figure}

The intensity profile given by Eq.~\eqref{LOSDM} can be used to find the mean number of photons per pixel, $x_\text{var,DM}^p$, integrated over the energy range from 1.9--11.9~GeV.  This is accomplished by convolving the energy-integrated intensity profile with the PSF  and multiplying by the exposure and pixel solid angle. Once the values of $x_\text{var,DM}^p$ are known, the dark-matter flux PDF easily follows from the generating function given by Eq.~\eqref{G_gen}.  The resulting flux PDF in the ROI is shown by the dashed-green line in Fig.~\ref{fig:PDF-no-diff}.

As a cross-check of this result, we also obtain the flux PDF by simulating a large number of data sets with Poisson counts determined by $x_\text{var,DM}^p$.  An example of a simulated map of counts is shown in the left panel of Fig.~\ref{fig:maps-no-diff}; its associated flux PDF is simply the histogram of the photon counts per pixel.  In this way, we build the flux PDFs for 500 individual data sets and then average them together to give the ``typical" simulated dark-matter flux PDF.  This simulated flux PDF is shown as the green histogram in Fig.~\ref{fig:PDF-no-diff} and matches well with the prediction from the generating-function approach, thereby validating our procedure for diffuse dark-matter emission.

\subsection{Point-source excess}
\label{sec:PS}
 
Next, we consider the possibility that the gamma-ray excess arises instead from point sources that have an energy spectrum compatible with that observed.  The flux PDF for the point sources depends on the generating function $P(t)$, which in turn depends on the mean number of $m$-photon sources per pixel, $x_m$.  To determine the $x_m$, one must know the source-count distribution $dN/dS$.  This follows from the spatial distribution $n(r)$ and luminosity function $dN/dL$ of the sources.  We therefore begin our calculation of the point-source flux PDF by motivating our assumptions for $n(r)$ and $dN/dL$.

To produce the spatial morphology of the excess, the intensity profile of the point-source emission should follow that in Eq.~\eqref{LOSDM}.  Therefore, we consider a population of sources with a normalized number-density profile
\es{nr}{
n(r) \propto r^{-\delta}
}
that is spherically symmetric about the GC.  The observed intensity profile is thus proportional to the LOS integral 
\es{LOS}{
 \int\! d \ell\, n[r(\ell, \psi)]   \propto \psi^{1 - \delta} \,.
}
Comparing with Eq.~\eqref{LOSDM}, we see that a spatial distribution with $\delta = 2 \gamma$ gives the desired intensity profile; when $\gamma = 1.26$, it follows that $\delta \approx 2.5$ and the intensity profile falls off roughly as $\psi^{-1.5}$.  
Intriguingly, the spatial distribution of LMXBs in M31 also appears to follow Eq.~\eqref{nr} with $\delta \approx 2.5 \pm 0.2$~\cite{astro-ph/0610649, 1207.6047}. 

With the spatial distribution fixed by the observed intensity profile of the excess, the point-source emission is then specified by the luminosity function.  We assume that the point sources have a luminosity function 
\es{LumFunc}{
\frac{d N}{d L} \propto L^{-\alpha_L}\,.
}
This choice is motivated by MSP luminosity functions considered in the literature, which are typically of this power-law form with indices $\alpha_L \sim 1$--$2$.  These luminosity functions are either derived from pulsar theory (see Ref.~\cite{1411.2980} for a discussion) or constructed~\cite{1407.5583} from the MSPs detected by \emph{Fermi}~\cite{1305.4385}.  However, there remains much uncertainty in both approaches.  For concreteness, we take $\alpha_L = 1.4$.  This is a relatively conservative choice, as testing the flux PDF on a point-source population with a more shallow luminosity function---\emph{i.e.}, one more heavily dominated by its brightest members and, hence, more obviously distinguishable from a diffuse source---could overstate its power.

The luminosity function for the point sources is truncated at both low and high $L$.  Because faint sources contribute little to the flux for the power law in Eq.~\eqref{LumFunc}, the flux PDF is not sensitive to the minimum luminosity cutoff; for concreteness, we choose $L_\text{min} = 2 \times 10^{33}$~ph~s$^{-1}$.  In contrast, the flux PDF is quite sensitive to the maximum luminosity cutoff; we therefore investigate a range of cutoffs $L_\text{max} = (0.75$--$20) \times 10^{36}$~ph~s$^{-1}$, which brackets possible values that have been inferred from \emph{Fermi} observations~\cite{1305.4385,1407.5583,1411.2980}.\footnote{Note that for sources with an energy spectrum similar to that of the excess, number luminosities in the energy range 1.9--11.9~GeV correspond to number luminosities roughly twice as large in the range $> 1$~GeV.  Therefore, this range of cutoffs roughly corresponds to $L_{\text{max}, > \text{1 GeV}} = (1.5$--$40) \times 10^{36}$~ph~s$^{-1}$, which includes values of $L_{\text{max}, > \text{1 GeV}}$ considered in Ref.~\cite{1411.2980}.}  The normalization of the luminosity function for each value of $L_\text{max}$ is then chosen to give the observed excess.

Incidentally, note that the maximum cutoff determines the number of sources that can be detected above threshold.  Indeed, for $L_\text{max}$ at the high end of the considered range, it is possible that the observed number of detected sources already constrains the MSP scenario, as argued by Refs.~\cite{1305.0830, 1406.2706, 1407.5625}.  However, the exact value of $L_\text{max}$ at which this occurs is difficult to determine, as it depends on both the detection threshold, which varies with Galactic latitude~\cite{1108.1435}, and the spatial distribution and luminosity function of the MSPs.  We also emphasize that taking $L_\text{max}$ to be the same as that observed for the nearby MSPs is a critical assumption in these arguments and is subject to uncertainties from both the small sample size of high-luminosity MSPs and large errors in MSP distances and luminosities~\cite{1208.3045,1407.6271,1408.0281}.  In any case, for all but the high end of the range of $L_\text{max}$ that we consider, it is likely that the total number of point sources that are over the detection threshold is less than or comparable to the number of unassociated detected sources in the Inner Galaxy.
 
\begin{figure}[tb]
\begin{center}
\includegraphics[trim=0in 0in 0in 0in, clip=true, width=0.6\textwidth]{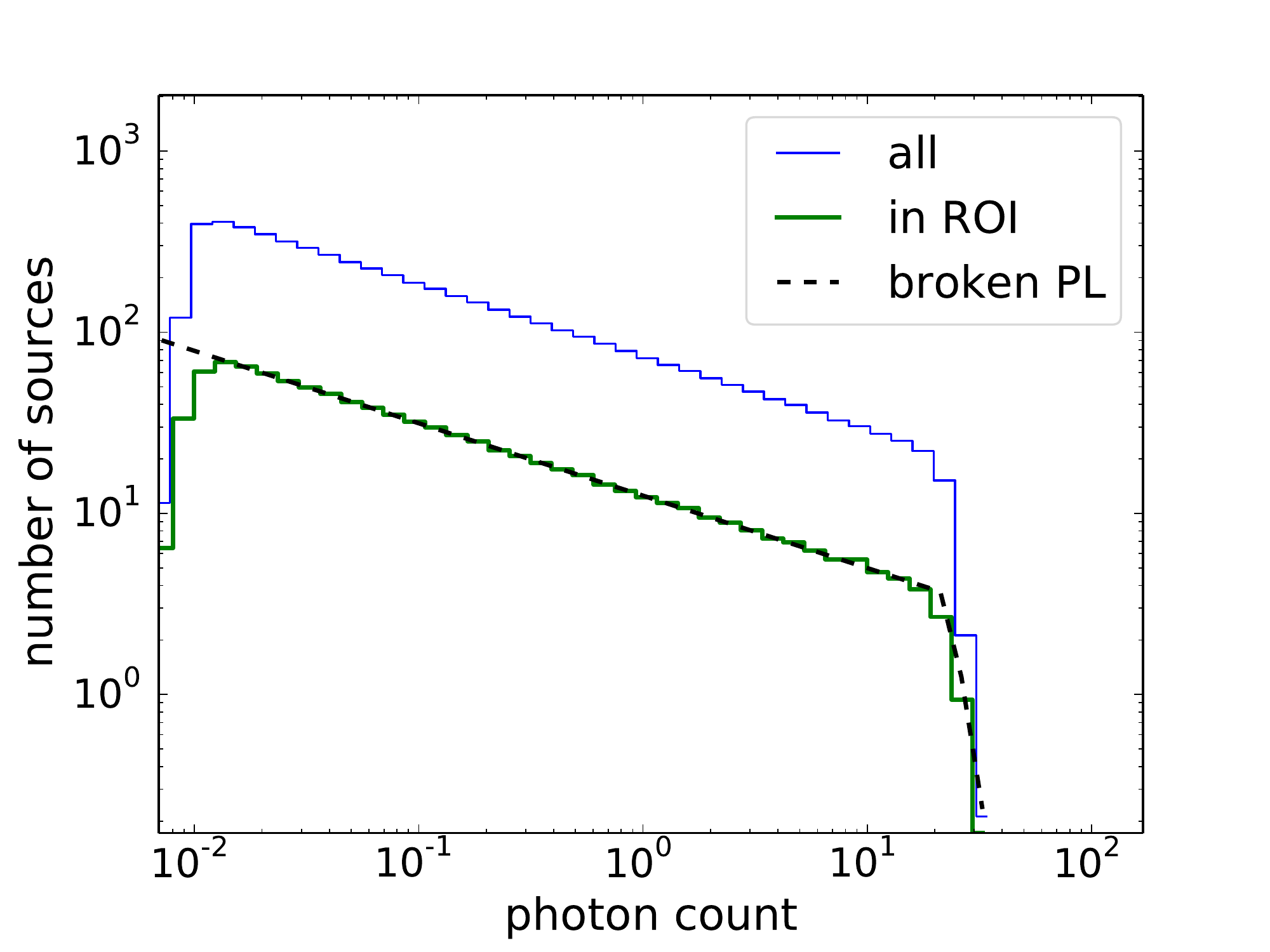}
\end{center}
\caption{The mean source-count distribution $dN/dS$ of MSP-like point sources in the Inner Galaxy, derived by averaging over 500 simulated data sets (blue histogram).  We assume a maximum luminosity cutoff $L_\text{max} = 5 \times 10^{36}$~ph~s$^{-1}$.  The corresponding distribution for the sources in the ROI is given by the green histogram, while the dashed-black line shows the best-fit broken power law parametrized by $\{A, S_b, n_1, n_2\}$.  The best-fit slope below the break is consistent with $n_2 \approx 1.4$, which follows the slope of the assumed MSP-like luminosity function.  The deficit of simulated sources at low counts relative to the fit results from  truncating the luminosity function at the faint end.  This discrepancy has little effect on the flux PDF because these faint sources contribute negligibly to the total point-source flux.}
\label{fig:dNdS}
\end{figure}

With the spatial distribution and luminosity functions set, we can determine the source-count distribution $dN/dS$.  If all of the point-sources were located exactly at the GC at a distance $r_0 = 8.5$~kpc away, then the standard expression for the flux (\emph{i.e.}, $L / 4 \pi r_0^2$) could be used to relate $dN/ dL$ to $dN / dS$.  However, the relation between the two functions is more subtle because of the finite spatial distribution $n(r)$ of the sources.  To proceed, we simulate 500 point-source realizations for a given $L_\text{max}$, calculate $dN / dS$ for each realization, and then find the average $dN/dS$.  The result is fit with a broken power law to give the best-fit parameters ${\alpha = \{A, S_b, n_1, n_2\}}$ (see Eq.~\eqref{dNdS}).  Fig.~\ref{fig:dNdS} shows the mean source-count distribution obtained for the benchmark value $L_\text{max} = 5 \times 10^{36}$~ph~s$^{-1}$.  The average simulated $dN/dS$ in the ROI is shown by the green histogram and is well approximated by the broken power-law fit (dashed-black line).  The blue histogram corresponds to the average simulated distribution for the Inner Galaxy.  

The generating function $P(t)$ for the point sources is then obtained using the mean number of $m$-photon sources per pixel, $x_m$, which is calculated from the source-count distribution as in Eq.~\eqref{xmPSF}.  The corresponding flux PDFs in the ROI, with and without the PSF, are given by the dotted-red and solid-orange lines in Fig.~\ref{fig:PDF-no-diff} for the benchmark $L_\text{max} = 5 \times 10^{36}$~ph~s$^{-1}$.  The results agree well with the histograms of the simulated PDFs.  To generate the latter, we simulate populations of point sources with the spatial distribution and luminosity function described above, calculate the corresponding flux as a function of position on the sky for a given exposure for each population, and generate maps of Poisson counts.  Examples of such maps, with and without the PSF, are shown in Fig.~\ref{fig:maps-no-diff} for one simulated realization.  The corresponding flux PDFs in the ROI are obtained by averaging the PDFs of 500 data sets simulated in this way.  The fact that the simulated flux PDFs match those obtained using the generating-function approach verifies that the procedure in Sec.~\ref{sec:photonstats}  correctly gives the flux PDF for a population of point sources.

Fig.~\ref{fig:PDF-no-diff} shows that the power-law tail at high counts clearly distinguishes the point-source flux PDFs from the nearly Poissonian dark-matter flux PDF, and that the effect of the PSF reduces the difference.  Furthermore, the point-source flux PDFs yield a larger fraction of empty or faint pixels, arising from the regions between sources.  Note, however, that these differences are appreciable for a given exposure only if the cutoff $L_\text{max}$ is sufficiently large, even in the case of zero background.  This is because the point-source flux PDF asymptotically approaches the dark-matter flux PDF as $L_\text{max}$ decreases and even the brightest point sources become relatively faint.\footnote{Put another way, from our earlier discussion of generating functions, and in particular, Eq.~\eqref{P_gen}, we saw that diffuse emission is indistinguishable from emission from a population of ``1-photon'' sources.  From Eq.~\eqref{xm}, we see that such a population occurs when: 1) the source-count distribution is close to or steeper than $dN/dS \propto S^{-2}$,  or 2) $S_b \lesssim 1$.  If either of these conditions are satisfied, then $x_1 \gg x_{m > 1}$ and all sources indeed contribute $\sim$1 photon.  In contrast, for the population that we consider, the slope of the source-count distribution is roughly that of the luminosity function (\emph{i.e.}, $dN/dS \propto S^{-1.4}$), which is a consequence of the fact that the sources are all located at similar distances $\sim$8.5~kpc away from us.  Thus, as long as $S_b$ (or equivalently, $L_\text{max}$) is sufficiently large, we will not be in the limit where the population of point sources is equivalent to a diffuse source.} Nevertheless, for sufficiently large $L_\text{max}$, it is exactly the differences in the flux PDFs that allow one to distinguish between the dark-matter and point-source scenarios.  

\subsection{Diffuse background}

\begin{figure}[p]
\makebox[\linewidth]{
\includegraphics[trim=.12in 0in .1in 0in, clip=true, height=0.26\textwidth]{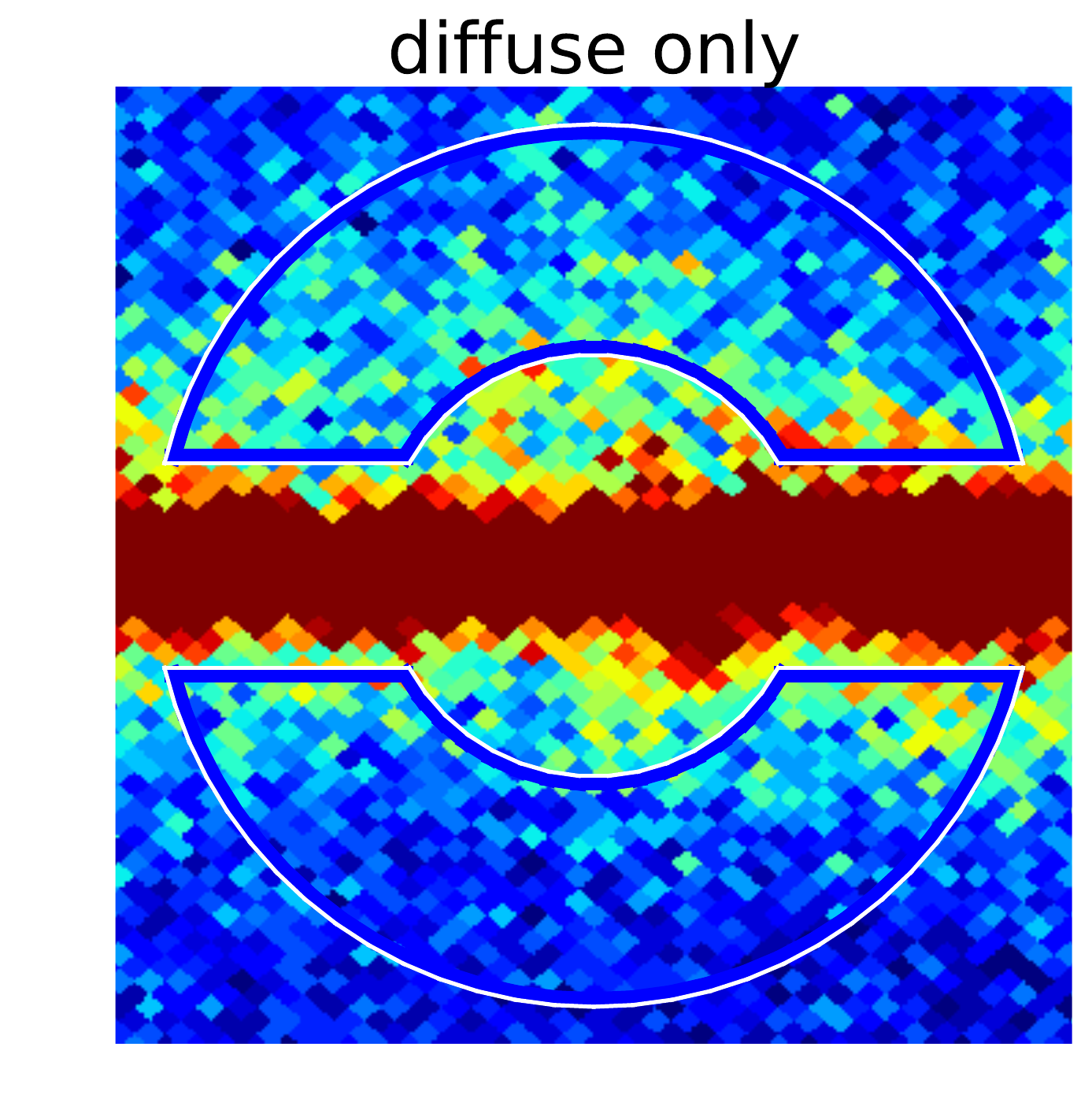} 
\includegraphics[trim=.12in 0in .1in 0in, clip=true, height=0.26\textwidth]{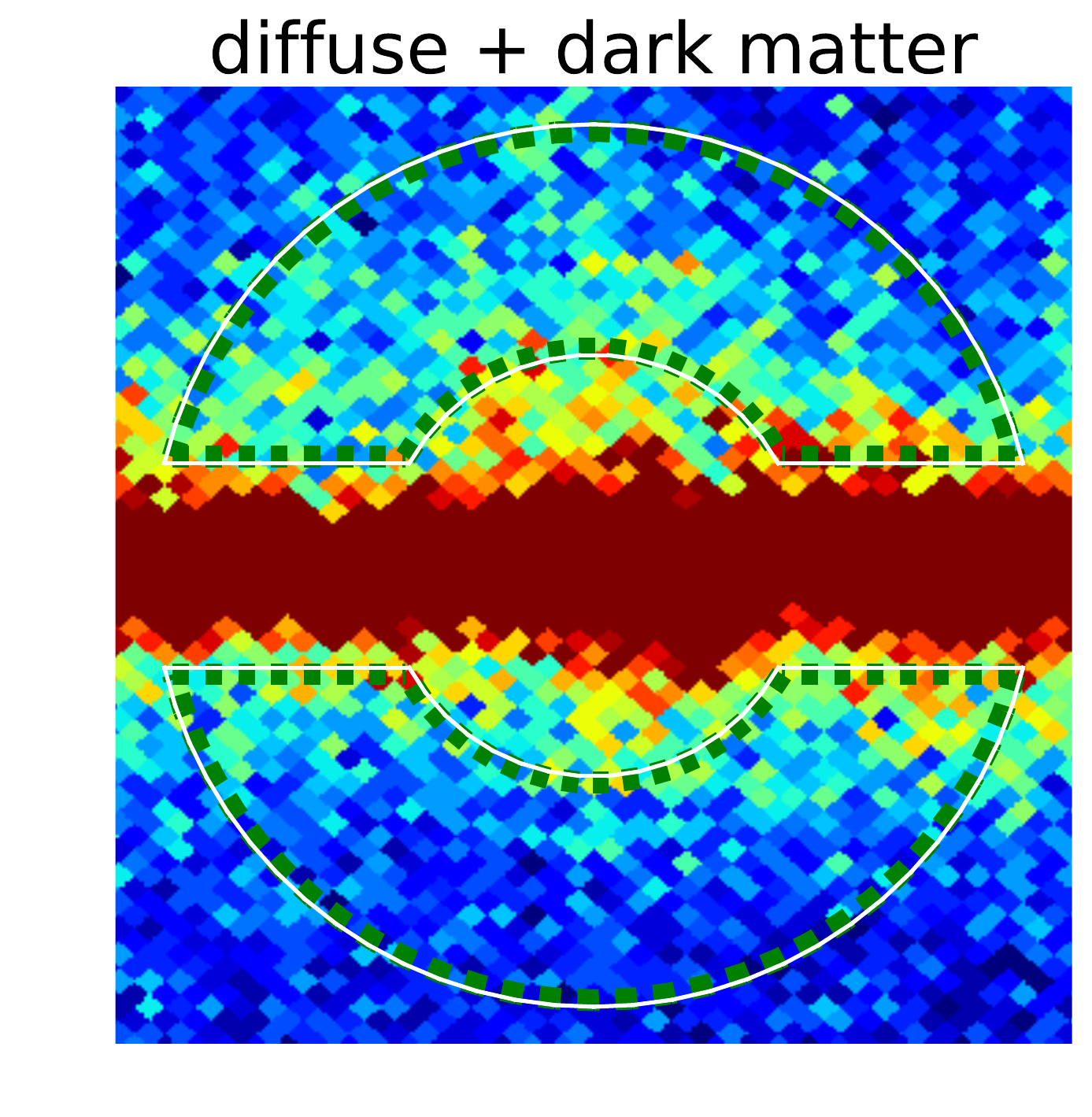}
\includegraphics[trim=.12in 0in .1in 0in, clip=true, height=0.26\textwidth]{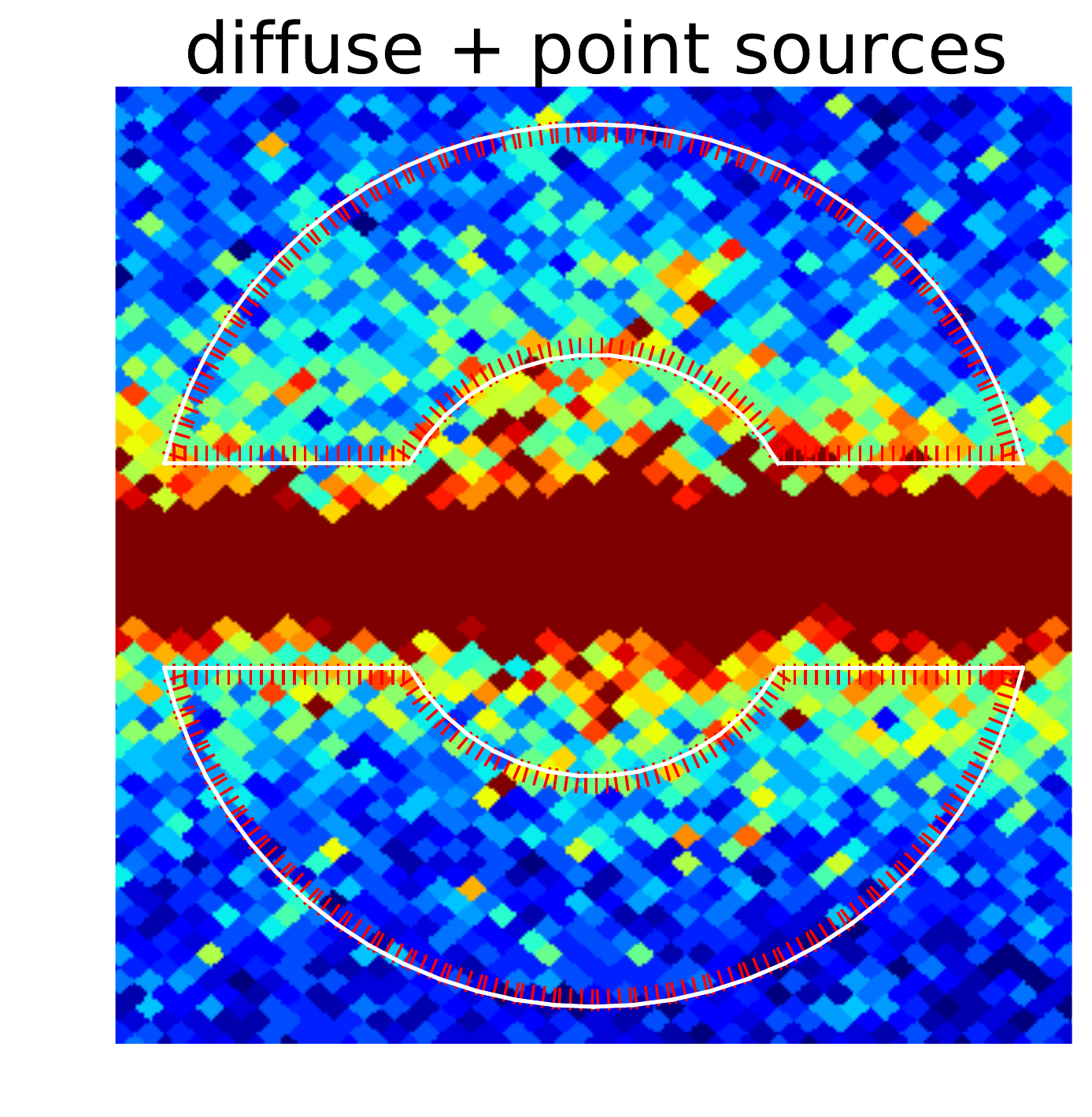}
\includegraphics[trim=.12in 0in .08in 0in, clip=true, height=0.26\textwidth]{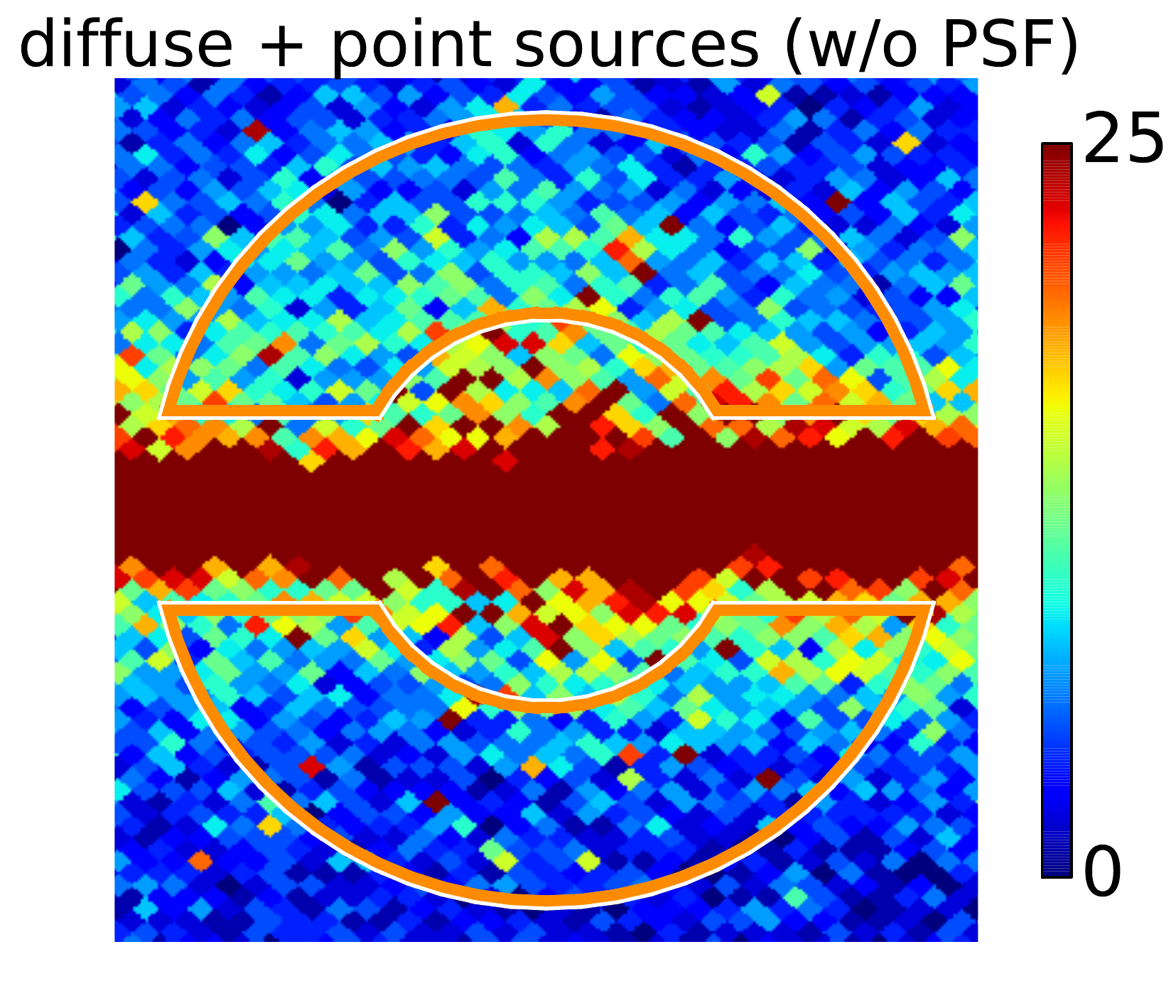} 
}
\caption{Maps of simulated photon counts in the Inner Galaxy.  The leftmost map shows counts (clipped at 25) from a \textsc{Galprop} model for the diffuse background, while the subsequent plots also include the excesses from Fig.~\ref{fig:maps-no-diff}. }
\label{fig:maps}
\vspace{0.25in}
\makebox[\linewidth]{
\includegraphics[trim=0in 0in 0in .5in, clip=true, width=0.6\textwidth]{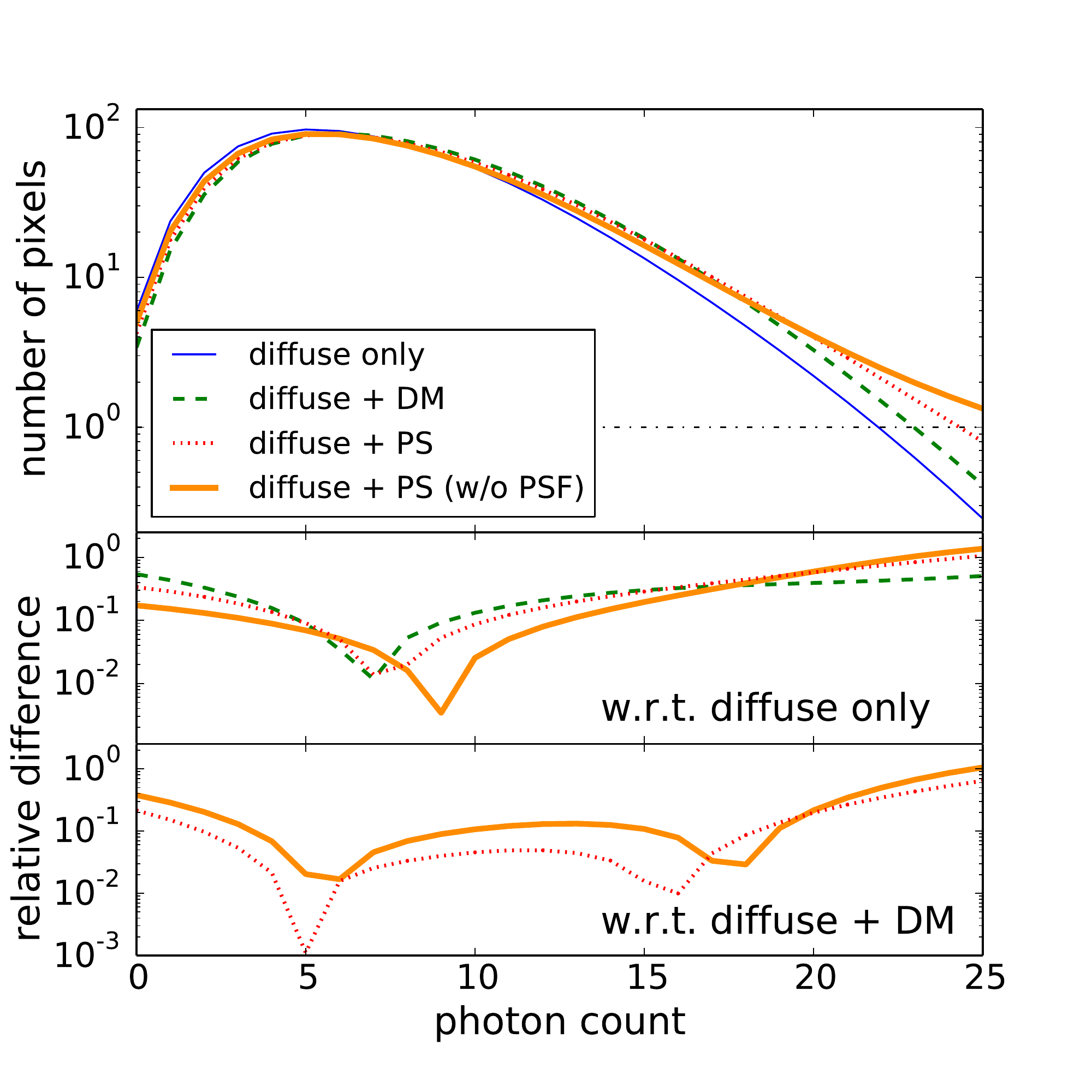}
}
\caption{\emph{(top)} Predicted flux PDFs in the ROI (normalized to the number of $nside = 128$ \textsc{HEALPix} pixels contained there) for each of the scenarios in Fig.~\ref{fig:maps}, calculated using the formalism and procedures described in Secs.~\ref{sec:photonstats}~and~\ref{sec:models}.  The solid-blue line shows the nearly Poissonian PDF of the \textsc{Galprop} model for the diffuse background.  Also shown are the combined PDFs for the diffuse background plus an excess arising from dark matter (dashed-green), point sources (dotted-red), and point sources without the PSF (solid-orange).  The dash-dotted--black line has the same meaning as in Fig.~\ref{fig:PDF-no-diff}.  \emph{(middle)} The relative difference of PDF $P_1$ with respect to PDF $P_2$ is defined as $2 |P_1 - P_2| / (P_1 + P_2)$.  Shown here are the relative differences of the dark-matter and point-source PDFs with respect to the PDF of the diffuse background.  \emph{(bottom)} The relative differences of the point-source PDFs with respect to the dark-matter PDF, which typically range from $\mathcal{O}(1-10)$\%, are plotted.  The differences increase at low photon counts due to the larger number of empty or faint pixels in the point-source scenarios and at high photon counts due to the power-law tails from individual bright sources.}
\label{fig:PDF}
\end{figure}

As Fig.~\ref{fig:PDF-no-diff} illustrates, the dark-matter and point-source flux PDFs can potentially be differentiated based on differences at the lowest and highest photon counts.  However, these differences may be obscured once the diffuse background model is properly accounted for.  The purpose of this section is to obtain the generating function for the Galactic diffuse background, determined from the mean photon counts per pixel, $x_\text{var,bg}^p$.  Eq.~\eqref{nuGen} can then be used to obtain the total flux PDF in the ROI for the dark-matter and point-source models of the excess.

We assume a diffuse-background model containing only a Galactic component, which is generated using the \textsc{Galprop} cosmic-ray--propagation code~\cite{astro-ph/9710124,astro-ph/9807150}.  The default parameters for the \textsc{Galprop} webrun routine~\cite{1008.3642} (including 3D diffusion) are used to generate differential intensity maps of the Galactic diffuse emission at the highest available resolution of $nside = 128$.  Maps for nine energies at regular logarithmic intervals in the range 1.9--11.9~GeV are generated and then interpolated and integrated to give maps of the total intensity in this range.  We use these intensity maps to find the mean diffuse-background counts per pixel, $x_\text{var,bg}^p$, following the same procedure used to construct $x_\text{var,DM}^p$.  It is straightforward to simulate maps of Poisson counts once the $x_\text{var,bg}^p$ are known.  One realization is shown in the left panel of Fig.~\ref{fig:maps}; the other panels in the figure additionally include the simulated excesses from Fig.~\ref{fig:maps-no-diff}.  The difficulty of distinguishing the dark-matter and point-source scenarios by eye highlights the need for using a statistic such as the flux PDF.

The mean counts per pixel, $x_\text{var,bg}^p$, can be substituted in Eq.~\eqref{G_gen} to find the expected diffuse-background flux PDF, which is plotted as the solid-blue line in Fig.~\ref{fig:PDF}.  The combined flux PDFs of the diffuse background with either dark matter or point sources are also shown.  As before, we check that the simulated flux PDFs in each scenario agree well with the predicted flux PDFs, although the corresponding histograms are not plotted.  Clearly, the presence of the dominating background obscures the differences between the dark-matter and point-source flux PDFs that were obvious in Fig.~\ref{fig:PDF-no-diff}.  As the middle subplot of  Fig.~\ref{fig:PDF} shows, the inclusion of the excesses only modifies the diffuse-background flux PDF at the $\mathcal{O}(1-10)$\% level.  Nevertheless, for sufficiently bright $L_\text{max}$, the flux PDFs of the dark-matter and point-source scenarios still differ from each other at a statistically discernible level, as is demonstrated by the bottom subplot of Fig.~\ref{fig:PDF}.  In the next section, we shall exploit this fact, using Bayesian model comparison to test the dark-matter and point-source hypotheses with the discriminating power of the flux PDF.

\section{Bayesian model comparison}
\label{sec:Bayesian}

Thus far, we have demonstrated how to calculate the flux PDF for dark-matter and point-source models of the excess.  We now investigate whether the observed flux PDF can provide evidence in favor of one of these two models, using the framework of Bayesian model comparison.  For a review of model comparison (and Bayesian methods, in general), see Ref.~\cite{0803.4089}; we simply provide a brief overview here.

In the context of Bayesian inference, the definition of a model (or a hypothesis) $\mathcal{M}$ must specify its parameters $\theta$, which range over the parameter space $\Omega_\mathcal{M}$, as well as the prior distributions $p(\theta | \mathcal{M})$ for these parameters.  Given a data set $d$, the Bayesian evidence for the model $\mathcal{M}$ is given by the prior-weighted average of the likelihood $p(d | \theta, \mathcal{M})$ as follows:
\es{evidence}{
p(d | \mathcal{M}) = \int_{\Omega_\mathcal{M}}\! d\theta\, p(d | \theta, \mathcal{M}) p(\theta | \mathcal{M})\,.
}
Thus, models with larger parameter spaces $\Omega_\mathcal{M}$ are penalized in the evidence; this disfavors overfitting the data, and may be seen as a formal expression of the principle of Occam's razor.

In our analysis, the data set $d$ is given by the observed flux PDF---\emph{i.e.}, the number of pixels $n_k$ in the ROI that contain $k$ photons.  The likelihood function is then
\es{L}{
p(d | \theta, \mathcal{M}) = \prod_{k = 0}^{k_\text{max}} {[p_k(\theta) \Np]^{n_k} \over n_k!} e^{- p_k(\theta) \Np } \,,
}
where $p_k$ is the predicted flux PDF and $\theta = \{x_\text{var}^p, \alpha\}$ specifies the model parameters.  Because the calculation of the evidence requires that the likelihood function be evaluated at many points in the parameter space, it is worthwhile to use recurrence relations and analytic results to speed up the calculation of the $p_k$.  We describe these methods in Appendix~\ref{App: speed}.  In addition, we employ the \textsc{MultiNest} package~\cite{0809.3437} to calculate the evidence.

In Eq.~\eqref{L}, only the pixels with photon counts below some $k_\text{max}$ are accounted for in the likelihood.  We choose $k_\text{max} = 25$, which is essentially the cutoff below which individual point sources are no longer resolved.\footnote{The cutoff should roughly be the total counts from the faintest detected source and the diffuse background in the pixel where the source is located.  The faintest detected source in the ROI contributes $\sim$20 photons in the relevant energy range~\cite{1108.1435}, and the \textsc{Galprop} model for the diffuse background yields an average of $\sim$7 photons per pixel in the ROI and energy range.}  Note that this is a conservative choice, as it only uses the information contained in the statistical fluctuations of the counts below the detection threshold, and ignores the extra information provided by detected sources.  In an analysis of the actual \emph{Fermi} data, choosing $k_\text{max}$ in this way will avoid contamination from bright sources that are members of other point-source populations.  We make this choice here as well, even though our simulated model of the gamma-ray sky does not contain such populations.

Given two models $\mathcal{M}_0$ and $\mathcal{M}_1$, the ratio of their evidences
\es{Bayes_factor}{
B_{10} = {p(d | \mathcal{M}_1) \over p(d | \mathcal{M}_0)}
}
is called the Bayes factor.  A Bayes factor $B_{10} > (<)~1$ indicates that an increase (decrease) of the belief in favor of $\mathcal{M}_1$ over $\mathcal{M}_0$ is supported by the data set $d$.  The strength of this belief is usually determined by interpreting the Bayes factor with respect to the empirically calibrated Jeffreys scale~\cite{Jeffreys61}; for example, $B_{10} \sim 10$ indicates strong evidence in favor of $\mathcal{M}_1$.  The Bayes factor is a useful statistic when the two models being compared do not share any parameters, as is the case for the dark-matter and point-source scenarios.

We use the Bayes factor to answer the following question: assuming the excess actually arises from a point-source population (with properties as described in  Sec.~\ref{sec:PS}), how strongly does the flux PDF of the counts below threshold provide evidence for the point-source hypothesis over the dark-matter hypothesis for a typical data set, and how does this depend on the luminosity cutoff $L_\text{max}$?  To answer this question, we examine cases in which the excess actually does arise from a point-source population with luminosity cutoff $L_\text{max}$ in the range (0.75--$20) \times 10^{36}$~ph~s$^{-1}$.  As an example of a ``typical" data set, we use the average flux PDF for a combination of the diffuse background and point sources with a given $L_\text{max}$, as calculated in Sec.~\ref{sec:models}.\footnote{In the context of frequentist hypothesis testing, this typical data set is usually referred to as the ``Asimov'' data set~\cite{1007.1727}.  It allows a formal approximation of both the median and the distribution of the statistical significance of a hypothesis, obviating the need to use Monte Carlo simulations to account for statistical fluctuations of the data.  Whether it can be used analogously in the context of Bayesian model comparison has not been formally shown, but we have applied our analyses to a small number ($\sim$15) of simulations for each of the considered values of $L_\text{max}$.  The resulting distribution of Bayes factors suggests that the Asimov data set indeed yields a good approximation of the median Bayes factor.}  This typical data set provides the $n_k$ needed in Eq.~\eqref{L}.

\begin{table}
\begin{center}
\begin{tabular}{|  c | c c |}
\hline
Parameter 	& Prior type & Prior range			 \Tstrut\Bstrut		\\   
\hline
$A_\text{DM}$ 	&  log-flat     & $[10^{-4}, 10^3]$  \Tstrut\Bstrut	\\
\hline
$A$ 		&  log-flat     	 & $[10^{-4}, 10^3]$~photons$^{-1}$ \Tstrut\\
$S_b$ 	&  linear-flat     & $[2, 100]$~photons \\
$n_1$ 	&  linear-flat     & $[2.05, 10]$ \\
$n_2$ 	&  linear-flat     & $[0.05, 1.85]$ \Bstrut\\
\hline
\end{tabular}
\end{center}
\caption{Prior ranges for the dark-matter and point-source models of the excess.  Identical dynamic ranges and priors for both of the normalization parameters $A_\text{DM}$ and $A$ are chosen, so that the overfitting factor in the comparison of the 1-parameter models is eliminated.  The values $S_b \approx 1$ and $n_2 \approx 2$, for which $x_1 \gg x_{m >1}$ and the point-source emission mimics that of a diffuse source, are avoided.  We also restrict $n_1 > 2$ to avoid divergences in the total flux inferred from the broken power-law model for $dN/dS$ and only consider values $n_2 > 0$ that yield falling luminosity functions.}
\label{tab:priors}
\end{table}

\begin{figure}[!t] 
\begin{center}
\includegraphics[trim=0in 0in 0in 0in, clip=true, width=0.8\textwidth]{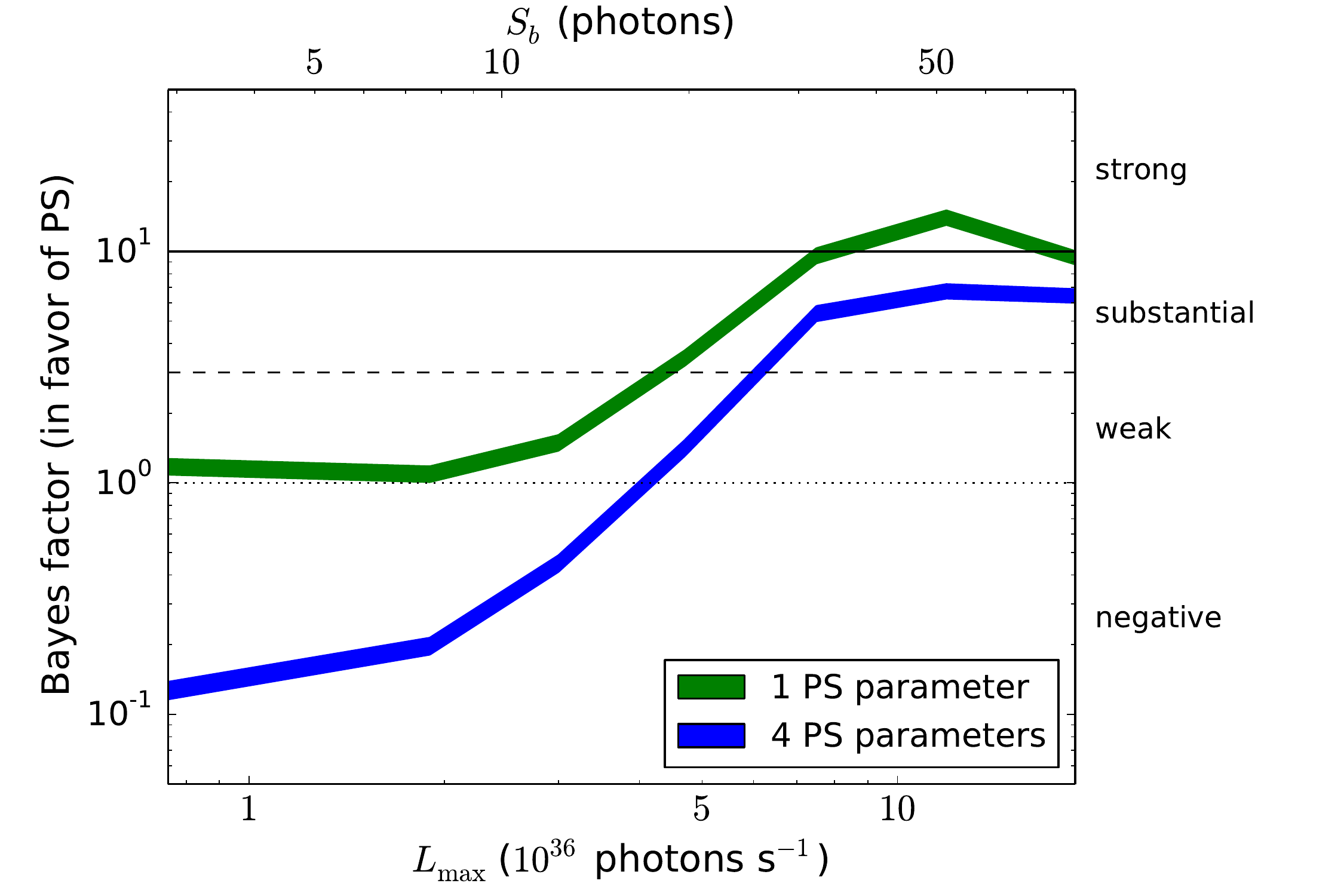}
\end{center}
\caption{The Bayes factor---the ratio of evidence for the point-source hypothesis to that for the dark-matter hypothesis---in the case that the excess arises from a population of point sources.  This is shown as a function of the maximum luminosity cutoff $L_\text{max}$ for the point-source population (or, alternatively, as a function of the corresponding source-count--distribution break $S_b$).  The blue band assumes a 4-parameter point-source model $\{A, S_b, n_1, n_2\}$, while the green band assumes a single-parameter model $\{A\}$; these point-source models are each compared to a single-parameter dark-matter model $\{A_\text{DM}\}$.  The width of each band denotes the $\pm1\sigma$ error on the numerical calculation of the Bayes factor, which is performed using \textsc{MultiNest}~\cite{0809.3437}.  The Jeffreys scale~\cite{Jeffreys61}, which measures the strength of the evidence in favor of the point-source hypothesis, is shown on the right axis.  For point-source populations with $L_\text{max} \lesssim (2$--$3) \times 10^{36}$~ph~s$^{-1}$, the flux PDF of the point sources is indistinguishable from that of dark matter, and hence the point-source models are not preferred.  In the case of the 4-parameter point-source model, the dark-matter model is actually preferred, as the former is penalized for having a larger number of parameters.  However, point-source models are typically preferred over the dark-matter model for point-source populations with $L_\text{max} \gtrsim6 \times 10^{36}$~ph~s$^{-1}$.  Ref.~\cite{1411.2980} suggests that a population of MSPs with such a cutoff could be responsible for the excess, while still being consistent with the observed number of point sources above the detection threshold.}
\label{fig:bayes-factor}
\end{figure}

Calculation of the Bayes factor requires the predicted flux PDFs $p_k$ for the diffuse plus dark-matter ($\mathcal{M}_0$) and the diffuse plus point-source ($\mathcal{M}_1$) models, obtained using the generating-function procedure.  For the dark-matter model, we parametrize the excess with a single normalization parameter $A_\text{DM}$ by writing $x_\text{var}^p = x_\text{var,bg}^p + A_\text{DM} x_\text{var,DM}^p$, where $x_\text{var,DM}^p$ is defined as in Sec.~\ref{sec:DM}; in this case, $\theta = \{A_\text{DM}\}$ in Eq.~\eqref{L}.  We consider two different point-source models: 1) a single-parameter model, with only $\theta = \{A\}$ allowed to vary and $\{S_b, n_1, n_2\}$ fixed to their true values (which, we recall from Sec.~\ref{sec:PS}, are determined by fitting the $dN/dS$ found from the average of many simulations with a broken power law, for given values of $L_\text{max}$), and 2) a 4-parameter model with $\theta = \{A, S_b, n_1, n_2\}$ all allowed to vary.  The extent of the parameter space $\Omega_\mathcal{M}$ for each model is determined by the choice of prior ranges, which are shown in Table~\ref{tab:priors}.

Fig.~\ref{fig:bayes-factor} shows the results of the model comparison.  The Bayes factor in favor of the point-source hypothesis is plotted as a function of $L_\text{max}$ for both sets of comparisons: 1) 1-parameter dark-matter model vs. 1-parameter point-source model (green), and 2) 1-parameter dark-matter model vs. 4-parameter point-source model (blue).  Substantial  evidence for the point-source models is found if the excess arises from a point-source population with a luminosity cutoff $L_\text{max} \gtrsim 6 \times 10^{36}$~ph~s$^{-1}$.  As concluded in Ref.~\cite{1411.2980}, such a population could plausibly comprise the excess without contributing too many detected sources.  Therefore, the flux PDF analysis can distinguish relevant point-source models from a dark-matter model of the excess.  

Also apparent in Fig.~\ref{fig:bayes-factor} is that the flux PDF of the point sources is indistinguishable from that of a dark-matter excess at $L_\text{max} \lesssim (2$--$3)\times 10^{36}$~ph~s$^{-1}$, and neither model is preferred if their respective parameter spaces and priors are identical.  Because the normalization parameters $A_\text{DM}$ and $A$ have the same prior constraints, the Bayes factor comparing the 1-parameter models is accordingly unity at low $L_\text{max}$.  In contrast, the Bayes factor comparing the 1-parameter dark-matter model and the 4-parameter point-source model asymptotes to a value less than unity that depends on the volume of the parameter space of the point-source model.  This results from the overfitting penalization of the 4-parameter point-source model with respect to the 1-parameter dark-matter model.  The interpretation is then that the dark-matter model is preferred at low $L_\text{max}$ due to its simplicity, even though the flux PDF does not otherwise discriminate between the models.

We emphasize that our calculations assume that the observed flux PDFs are given by the typical data set for each value of $L_\text{max}$.  For actual realizations of the data, which are subject to Poisson fluctuations, there will be scatter in the Bayes factor~\cite{1012.3195,1101.4822} around the curves in Fig.~\ref{fig:bayes-factor}.  Preliminary calculations based on a small number ($\sim$15) of simulated realizations for each value of $L_\text{max}$ show that this scatter can become quite large (spanning more than a decade) once the Bayes factor is significant enough to provide substantial evidence.  The scatter can be quantified more precisely by performing a comprehensive calculation of the distribution of Bayes factors as a function of $L_\text{max}$, which requires model comparisons for a large number of simulated data sets at each value of $L_\text{max}$.  Such a calculation is computationally intensive, but necessary for understanding the rate of false-positive and false-negative errors for a flux-PDF--based model comparison.  For example, if only a weak Bayes factor is obtained from the real data, one would need to quantify the probability that point sources with a bright $L_\text{max}$ are still present even though the flux PDF, due to realization noise, happened to be nondiscerning.\footnote{In this case, increasing $k_\text{max}$ or examining properties of the detected sources might also provide additional information.  Other Bayesian diagnostics, such as the complexity~\cite{RSSB:RSSB353} or various information criteria~\cite{astro-ph/0401198, astro-ph/0610126,0803.4089}, could also prove useful.} We leave this to future work.

\section{Discussion}
\label{sec:discussion}

It is possible that the excess of GeV gamma-ray photons observed at the GC and the Inner Galaxy arises from dark-matter annihilation.  However, before making such a strong claim, other alternative explanations must be ruled out by the available data.  Here, we have focused on the possibility that the excess arises from a population of unresolved point sources or structures, which may include millisecond pulsars or other astrophysical objects.  The properties of such a population can be fully specified by its spectrum, spatial distribution, and luminosity function, and many of these properties will necessarily be fixed if the population is indeed responsible for the excess.   Thus, the relative simplicity of the point-source hypothesis, as compared to other astrophysical explanations for the excess, suggests that it may be the easiest and cleanest to test.

Working in the framework of Bayesian model comparison, we have demonstrated that the statistics of the excess photons---specifically, the flux PDF---can distinguish point sources from dark matter, if the diffuse background is well determined.  For a \textsc{Galprop}-generated diffuse background and a typical data set, this is the case if the brightest members of the population have a luminosity $L_\text{max} \gtrsim 6 \times 10^{36}$~ph~s$^{-1}$ in the energy range 1.9--11.9~GeV.  Our results show that the statistics of the photon counts below threshold contain valuable information that is complementary to the number of detected sources above threshold.

Nevertheless, just as any claim of an excess can only be made with respect to a specified background model, any evidence for the presence of point sources provided by the statistical methods we propose depends on this  model.  Determining the diffuse-background model (\emph{i.e.}, the mean counts $x_\text{var,bg}^p$) from the noisy data is a difficult problem that is closely related to both the flux PDF and the detection of point sources.  The \emph{Fermi} Collaboration builds its diffuse model~\cite{1202.4039}, which is intended to be used primarily for background subtraction in the study of point sources, by fitting a combination of \textsc{Galprop}-generated maps (which give the inverse-Compton emission) and gas--column-density maps (which give the bremsstrahlung and $\pi^0$-decay emission) to the data.  Alternatively, for the statistical analyses of the photon counts we propose, it may be possible to approach this problem in a more purely data-driven manner.  That is, we may be able to delineate diffuse and point-source components of the data using procedures based on statistical and spatial criteria---without heavily relying on input from cosmic-ray--propagation models or detailed knowledge of other uncertain astrophysical quantities.  We have conducted preliminary investigations of whether wavelet-based procedures~\cite{1003.5613,1206.2787} for building the diffuse model from noisy data are sufficient for our flux-PDF method, with encouraging results.  Other data-driven algorithms~\cite{2013arXiv1311.1888S, 1410.4562} for reconstructing the diffuse emission may also prove useful.

Furthermore, in this study, we have assumed relatively simple models for both the diffuse background and the excess.  In particular, we only include Galactic emission (estimated using \textsc{Galprop}) in our background model.  However, the inclusion of sub-dominant components of the background, including other point-source populations, may complicate matters.  In this case, sideband analyses---of regions of sky other than the Inner Galaxy or at energies outside of the peak of the excess---may prove necessary to understand how these additional background components contribute to the flux PDF.  All of these caveats should and will be carefully taken into account in the application of these methods to the real data, which is work in progress.

On the other hand, we also note that the flux-PDF analyses presented here are relatively simple and may understate the potential for discriminating between point sources and dark matter.  For example, we focus only on a particular (and somewhat arbitrarily chosen) ROI and also neglect the energy dependence of the photon counts.  It is possible that analyses that account for the spatial and energy dependence of the flux PDF could prove even more powerful.  Combining the flux PDF with other statistics, including two-point statistics such as the angular autocorrelation function or the angular power spectrum, might further improve our ability to discern the presence of unresolved point sources.  Such two-point statistics would quantify the clustering in the photon counts, which is apparent in the point-source excess shown in Fig.~\ref{fig:PSF-maps}.  For pulsating point sources, even the statistics of the photon arrival times may be informative~\cite{1106.4813}.

Additionally, both the detection threshold for resolving point sources and the complementary ability of the flux PDF to probe unresolved sources improve along with the PSF.  In this work, we have assumed a relatively narrow PSF, which can be achieved by restricting the analysis to a class of events with a cut on the event-reconstruction parameter CTBCORE~\cite{1406.0507}.  The upcoming \emph{Fermi} Pass 8 data set, which will be the product of a systematic and comprehensive revision of the entire \emph{Fermi} event-level analysis, is expected to further improve on this PSF~\cite{1303.3514}.  This gives reason to be optimistic about whether the question of the point-source contribution to the excess will ultimately be settled with \emph{Fermi} data alone.  However, it is possible that we will have to wait until future instruments with improved angular resolution---such as the confirmed GAMMA-400  telescope (which will be optimized for energies $\sim$100~GeV)~\cite{1412.4239} and the proposed PANGU telescope (which will target the $\sim$0.01--1~GeV range)~\cite{1407.0710}---can weigh in on the matter.

Finally, although understanding the photon statistics of point sources is of immediate interest for the issue of the excess in the Inner Galaxy, it may eventually be necessary for the interpretation of searches for annihilation in dwarf galaxies and extragalactic dark-matter halos at higher Galactic latitudes. Extragalactic point sources---such as active galactic nuclei and star-forming galaxies---comprise a guaranteed background for these observations.  Thus, by developing our understanding of these statistics, we not only place ourselves in a good position to begin solving the mystery of the GeV excess, but we also prepare ourselves for future observations of the gamma-ray sky.

\section*{Acknowledgments}
We thank Shin'ichiro Ando, Ewan Cameron, Kyle Cranmer, Douglas Finkbeiner, Vera Gluscevic, Marc Kamionkowski, Samuel McDermott, Jennifer Siegal-Gaskins, Joseph Silk, Tracy Slatyer, Meng Su, Jesse Thaler, Neal Weiner, and Wei Xue for helpful discussions, as well as Stephen Portillo and Jeffrey Santner for technical assistance.
S.K.L. and B.R.S. respectively thank the MIT Center for Theoretical Physics and Princeton University for hospitality during this work.  B.R.S was supported in part by a Pappalardo Fellowship in Physics at MIT and in part by the US Department of Energy under grant Contract Number DE-SC00012567. 

\appendix

\section{Methods for calculating the $p_k$} 
\label{App: speed}

In this Appendix, we describe the numerical calculation of the $p_k$.  While it is straightforward to use the formalism in Sec.~\ref{sec:photonstats} to calculate these quantities, it is necessary to perform the calculations quickly in practice.  The reason is that the probabilities $p_k$ must be calculated multiple times as the likelihood function is scanned over by varying the parameters $\alpha$. 

We proceed by writing Eq.~\eqref{nuGen} as
\es{pkA}{
p_k = {1 \over k!} {d^k \over dt^k} \left[ \sum_{p=1}^{N_\text{pix}} e^{f_p(t) }\right]_{t=0} \,,
}
where 
\es{}{
f_p(t) = x_\text{var}^p (t-1) + \sum_{m=1}^\infty x_m (t^m - 1) \,.
}
The expression for the $x_m$ is given explicitly in Eq.~\eqref{xmA}, which may be used along with properties of the incomplete gamma functions to compute the $x_m$ recursively.

Because differentiating is slow numerically, we take the derivatives with respect to $t$ in Eq.~\eqref{pkA} analytically.  Towards that end, we need to evaluate the derivatives 
\es{fpk}{
f_p^{(k)} &\equiv \left. {d^k \over dt^k} f_p(t) \right|_{t=0} \\
&= \left\{ 
\begin{array}{ll}
-(x_\text{var}^p + \sum_{m=1}^\infty x_m) \,, & k = 0 \,, \\
x_\text{var}^p + x_1 \,, & k = 1 \,, \\
k! \,x_k \,, & k > 1 \,.
\end{array} \right.
}
In this work, we consider cases where $n_1 > 2$ and $n_2 < 2$, allowing us to perform the sum over $m$ appearing above analytically:
\es{}{
\sum_{m=1}^\infty x_m = \frac{A}{\Np} \left[ {S_b^{1 - n_1} \over n_1-1}  + {S_b^{1 - n_2} \over n_2-1} - \Gamma(1- n_1, S_b) - \Gamma(1 - n_2) + \Gamma(1 - n_2, S_b) \right] \,.
}
  It then follows that 
\es{}{
p_k = {1 \over k!} \sum_{p=1}^{N_\text{pix}} F_p^{(k)} \,,
}
with
\es{}{
F_p^{(k)} \equiv \sum_{n=0}^{k-1} 
\left( \begin{array}{c} k-1 \\ n \end{array} \right)
f_p^{(k-n)} f_p^{(n)} \,, \qquad k \geq 1 \,.
}
This last expression provides a recurrence relation that can be used to compute the $F_p^{(k)}$---and hence, the $p_k$---iteratively.

Another method for cutting down the computation time comes from noting that the pixel dependence of the $F_p^{(k)}$ arises solely from the spatially varying diffuse emission.  That is, two pixels with the same value of $x_\text{var}^p$ will have the same $F_p^{(k)}$.  If the $x_\text{var}^p$ lie in the range $[0, c_\text{max} ]$ over all unmasked pixels, we may then write 
\es{pkNPC}{
p_k = {1 \over k!} \sum_{c=0}^{c_\text{max}} N_p(c) \left. F_p^{(k)} \right|_{x_\text{var}^p = c} \,.
}
Above, $N_p(c)$ is the number of unmasked pixels where $x_\text{var}^p = c$.  Evaluating the $p_k$ using Eq.~\eqref{pkNPC} substantially cuts down on the number of times that the functions $F_p^{(k)}$ must be evaluated.   

\bibliographystyle{JHEP}
\bibliography{arXiv_v2}

\end{document}